\documentclass[review,12pt]{elsarticle}

\usepackage{amssymb}
\usepackage{xcolor}
\usepackage{ulem}
\usepackage{lineno}

\journal{Carbon}

\begin{document}

\begin{frontmatter}

\title{Assessing the reliability of the Raman peak counting method for the characterization of SWCNT diameter distributions: a cross-characterization with TEM}

\author[LEM]{Alice Castan\corref{cor1}}
\cortext[cor1]{Corresponding authors. \\E-mail: alicecastan1@gmail.com, sofie.cambre@uantwerpen.be, annick.loiseau@onera.fr}
\author[ECMPL]{Salom\'e Forel}
\author[LEM]{Fr\'ed\'eric Fossard}
\author[ECMPL]{Joeri Defillet}
\author[LEM]{Ahmed Ghedjatti}
\author[ECMPL]{Dmitry Levshov}
\author[ECMPL]{Wim Wenseleers}
\author[ECMPL]{Sofie Cambr\'e\corref{cor1}}
\author[LEM]{Annick Loiseau\corref{cor1}}
\address[LEM]{Laboratoire d'Etude des Microstructures, CNRS-ONERA, University Paris-Saclay, Ch\^{a}tillon, France}
\address[ECMPL]{Experimental Condensed Matter Physics Laboratory, University of Antwerp, Belgium}

\address{}

\begin{abstract}
Resonant Raman spectroscopy is a widely used technique for single-walled carbon nanotube (SWCNT) characterization, in particular in the radial breathing mode (RBM) range which provides direct information on the structure of the nanotube in resonance. The RBM peak counting method, \textit{i.e.} acquiring Raman spectrum grids on a substrate with a select set of discrete laser lines and counting RBM peaks as single nanotubes, is frequently used to characterize SWCNT growth samples, despite the many factors that can induce errors in the results. In this work, we cross-characterize the diameter distributions obtained through this methodology with diameter distributions obtained by counting SWCNT diameters in transmission electron microscopy (TEM) and discuss the different results and biases between the techniques. This study is performed on a broad diameter distribution sample, and on two chirality-enriched samples whose chirality distributions are determined by photoluminescence excitation spectroscopy (PLE) and statistical analysis of high resolution TEM (HRTEM) images. We show that the largest differences between the Raman peak counting and TEM diameter distributions stem from the chirality-dependence of SWCNT Raman cross-sections and the patchy vision offered by the use of only a few discrete excitation wavelengths. The effect of the substrate and TEM-related biases are also discussed.
\end{abstract}


\end{frontmatter}

%% \linenumbers

%% main text
\newpage

\section{Introduction}
\label{Introduction}
Since the properties of single-walled carbon nanotubes (SWCNTs) are determined by their atomic structure \cite{saito1992}, exploiting these properties within the scope of a specific application ideally requires having access to chirality-pure, or chirality-enriched SWCNT samples. Over the past two decades, a tremendous effort has been made to produce such samples by selective growth using chemical vapor deposition (CVD) \cite{jourdain2013current, liu2016controlled, he2019designing} or by post-synthesis sorting \cite{hersam2008progress, fagan2019aqueous}, leading to an increasing control over sample composition, and even claims of close to single-chirality samples. However, an often overlooked aspect of this research field is the significant difficulty related to the assessment of sample composition \cite{li2019preparation, yang2017bilayer, tian2015reference}.\\\
Resonant Raman spectroscopy (RRS) is probably the most commonly used technique for growth selectivity assessment on flat substrates \cite{yang2020chirality}. By looking at the Raman-active radial breathing modes (RBM) of the SWCNTs as a function of laser excitation wavelength, we can have access to the diameter, semiconducting or metallic character, and chirality \cite{dresselhaus2005raman} of the probed SWCNTs. Because it gives access to so much information, Raman spectroscopy of SWCNTs has been extensively studied, and multiple groups have come up with methodologies to assess the composition of a given sample. Studies have been dedicated to the development of reliable Raman-based methodologies to determine the chirality of individual SWCNTs \cite{levshov2017accurate}, or the composition of bulk SWCNT samples \cite{jorio2005quantifying, araujo2010resonance}. Other methods have been developed where Raman spectroscopy data is cross-correlated with data from other techniques like, for instance, photoluminescence excitation (PLE) spectroscopy \cite{jorio2006carbon} or transmission electron microscopy (TEM) \cite{pesce2010calibrating}.  However, all of these methods either cannot be realistically adapted to a large ensemble of nanotubes on a substrate \cite{levshov2017accurate}, or rely on the use of multiple widely-tunable laser systems that are only available in a few laboratories worldwide, and the experiments and data treatment times are too long to be performed as a routine sample analysis \cite{jorio2005quantifying, araujo2010resonance, jorio2006carbon, pesce2010calibrating}.\\\
Because these methods were not easy to implement as a routine evaluation of a large number of samples, many research groups have shifted towards using variations of the same micro-Raman mapping-based alternative methodology. This methodology relies on accumulating spectra at the nodes of a grid over the surface of the sample in a Raman microscopy setup using several discrete laser lines, and counting RBM peaks as individual SWCNTs regardless of their intensity \cite{zhang2015n, yang2015growing, an2016chirality, zhao2016chemical, he2012diameter, castan2017new, zhang2016growth, qin2014growth, zhang2015diameter, cheng2018selective, zhang2019growth}. We will refer to this as the Raman peak counting, or RBM peak counting methodology. Though widely employed in recent years, the accuracy of this specific way of using Raman spectroscopy has not been verified through cross-characterization with other techniques and theoretical calculations of the Raman cross-sections. \\\
TEM can be used to measure SWCNT diameters \cite{fleurier2009transmission}, or assign chiralities using high-resolution TEM (HRTEM) \cite{ghedjatti2017structural}, or electron diffraction (ED) \cite{jiang2006robust, allen2011review}. Comparison of statistical data obtained by TEM and RBM peak counting for selectivity assessment is very rarely seen in the literature, in part because acquiring and analyzing TEM data can also be time consuming, and experimentally onerous. He \textit{et al.} reported differences between mean diameters extracted from TEM and Raman peak counting distributions, but explained the differences by their use of only one laser line for Raman characterization, leading to a bias in the detection of SWCNT populations \cite{he2012diameter}. Tian and coworkers recently showed, using four laser lines, that the RBM peak counting method gave erroneous results when looking at the semiconducting-to-metallic ratio in a SWCNT sample when comparing it to results obtained from ED \cite{tian2018validity}. These studies indicate that the impression of bypassing certain well-known factors such as chirality-dependent Raman cross-sections that are expected to affect the obtained sample composition when using the RBM peak counting method is misleading. The extent of the effect of such factors on the diameter distribution on SWCNT samples obtained by RBM peak counting has not been shown. In this work, we therefore assess the validity of the RBM peak counting method for determining diameter distributions.\\\
We propose to evaluate its accuracy for the assessment of the diameter distribution of a sample in two studies. We perform RBM peak counting with four excitation wavelengths to obtain an RBM peak frequency distribution of the samples, which gives a diameter distribution through the use of an empirical law, and compare the results to a TEM-extracted diameter distribution of the same samples. The first study is focused on a broad-diameter distribution sample meant to mimic a typical CVD-grown sample, containing isolated SWCNTs that were purified by density gradient ultracentrifugation (DGU) to check for discrepancies between the techniques. In the second study, we attempt to further understand the reasons for observed discrepancies by characterizing chirality-enriched samples sorted by aqueous two-phase extraction \cite{subbaiyan2014role, fagan2019aqueous} with much narrower diameter and chirality distributions. These samples are cross-characterized not only with Raman peak counting and TEM, but their composition is also assessed by coupling PLE, and statistical analysis of HRTEM images. 

\section{Results and discussion}

\subsection{Characterizing a SWCNT sample with a broad diameter distribution}
Potential sources of bias can arise from both the RBM peak counting method and TEM for the characterization of a SWCNT sample. In case of the Raman-based methodology, the use of an unadapted RBM frequency-to-diameter law, which depends significantly on environmental effects \cite{araujo2008nature} can lead to errors in the diameter measurements. Additionally, the chirality-dependence of Raman cross-sections, both predicted by theory \cite{popov2004resonant, popov2006resonant, jiang2005electron, machon2005strength, sato2010excitonic} and shown experimentally \cite{maultzsch2005radial, jungen2007raman, doorn2004resonant, piao2016intensity}, as well as the use of only a select number of excitation wavelengths, can lead to the over- or underestimation of certain SWCNT populations. Furthermore, Raman peak counting is typically performed on Si/SiO$_2$ substrates, for which it is known that interference effects can cause severe suppression or enhancement of the Raman signals at specific wavelengths depending on the oxide thickness \cite{monniello2019comprehensive, yoon2009interference, wang2008interference}. Regarding TEM, even though it theoretically allows the visualization, and characterization, of all nanotubes regardless of diameter, chirality, or metallicity, the degradation of SWCNTs under electron beam irradiation \cite{smith2001electron} primarily affects SWCNTs with small diameters, whose structures are more strained than their larger counterparts  \cite{warner2009investigating}. This can lead to an underestimation of the SWCNTs with the smallest diameters in the distribution. Moreover, since larger objects generate more contrast, they are more easily spotted by the TEM operator, which could again lead to a bias towards larger-diameter SWCNTs.
\begin{figure*}[h!]
	\centering
	\includegraphics[scale=0.24]{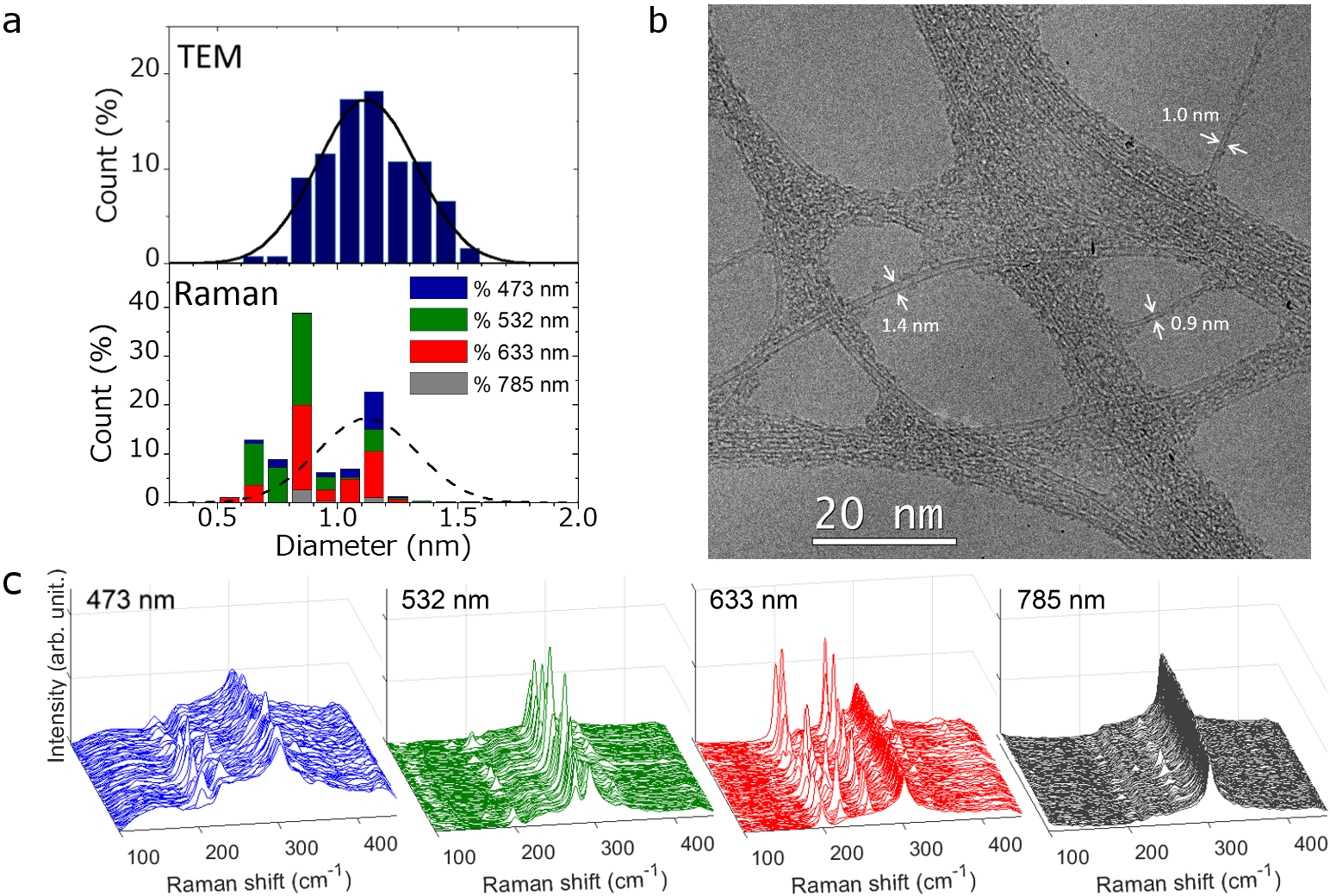}
	\caption{TEM and Raman characterizations of Sample 1: a) Top panel: TEM-extracted diameter distribution histogram, and Gaussian fit of the distribution (black line), Bottom panel: diameter distribution histogram from RBM peak counting showing the contribution of each excitation wavelength to each bin (blue, green, red, and grey columns), compared to the Gaussian fit of the TEM-extracted diameter distribution (black dashed line), b) typical TEM image of SWCNTs from Sample 1, and c) Raman spectra for one mapping experiment in the RBM spectral range for each of the 4 excitation wavelengths. The peak at 303 cm$^{-1}$ originates from the SiO$_2$ substrate.} 
	\label{dt-sorted-TEM-Raman}
\end{figure*}

In order to investigate the effects of all these aspects of both Raman peak counting and TEM on the diameter distribution assessment of a SWCNT sample, we first compared the two techniques for individualized SWCNTs obtained from the DGU purification of an as-synthesized SWCNT sample (see Methods section for details; further denoted as Sample 1). This sample was made to mimic a typical CVD-grown sample as studied in the literature. The same SWCNT dispersion was deposited both on a SiO$_2$/Si wafer with a 300 nm thermal oxide layer (which is the most occurring substrate for RBM peak counting in the literature \cite{zhang2015n, an2016chirality, castan2017new, zhang2016growth,  zhang2015diameter, cheng2018selective}) for Raman characterization (see Figure S2 in the electronic supplementary information (ESI) for a scanning electron microscopy (SEM) image) and a TEM grid for TEM observation. The diameter distributions were extracted from both techniques using the methodologies detailed in the Methods section. Figure \ref{dt-sorted-TEM-Raman} gives a summary of the Raman peak counting and TEM characterizations. The diameters were measured by TEM on 121 individually suspended SWCNTs (see Figure \ref{dt-sorted-TEM-Raman}.b for examples), resulting in a Gaussian distribution centered around 1.12 nm, with a full width at half maximum (FWHM) of 0.48 nm (see Figure \ref{dt-sorted-TEM-Raman}.a, and Figure S1 in the ESI). It should be noted that this diameter distribution contains a large percentage of sub-nanometer diameter SWCNTs, indicating that TEM in principle can detect these small diameters as well, even though they are more prone to beam damage. However, the fact that the leftmost two bins (diameters below 0.8 nm) have very low counts might suggest there is a bias at the very smallest diameters. We minimize the operator bias towards larger diameters by always proceeding with the same methodology: when an ensemble of nanotubes is spotted at low magnification, images of the entire zone are taken at a magnification high enough to resolve sub-nanometer nanotubes, and all suspended SWCNTs seen in the resulting images are counted.\\\
The Raman peak counting diameter distribution was extracted from 520 RBM peaks in a total of 484 spectra, counted on mappings at each excitation wavelength, examples of which are given in Figure \ref{dt-sorted-TEM-Raman}.c. Figure \ref{dt-sorted-TEM-Raman}.a compares TEM- and Raman-extracted diameter distributions. We can see that their shapes differ significantly: while the TEM distribution can be fitted by a Gaussian function except for a deviation for diameters below 0.8 nm, the Raman distribution is bimodal. We can notice two predominant populations, between 0.6 and 0.9 nm, and between 1.1 and 1.2 nm, and two gaps in the distribution between 0.9 and 1.1 nm, and between 1.2 and 1.6 nm. Additionally, about 40\% of the detected nanotubes have diameters between 0.8 and 0.9 nm, compared to less than 10\% in the case of the TEM-extracted distribution. This result shows the crucial importance of cross-characterization: the differences observed in the diameter distributions are significant, and would inevitably lead to even more substantial errors when using the Raman methodology to go further and determine the semiconducting-to-metallic ratio \cite{tian2018validity}, or the chirality distribution of the sample.\\\
\begin{figure*}[h!]
	\centering
	\includegraphics[scale=0.23]{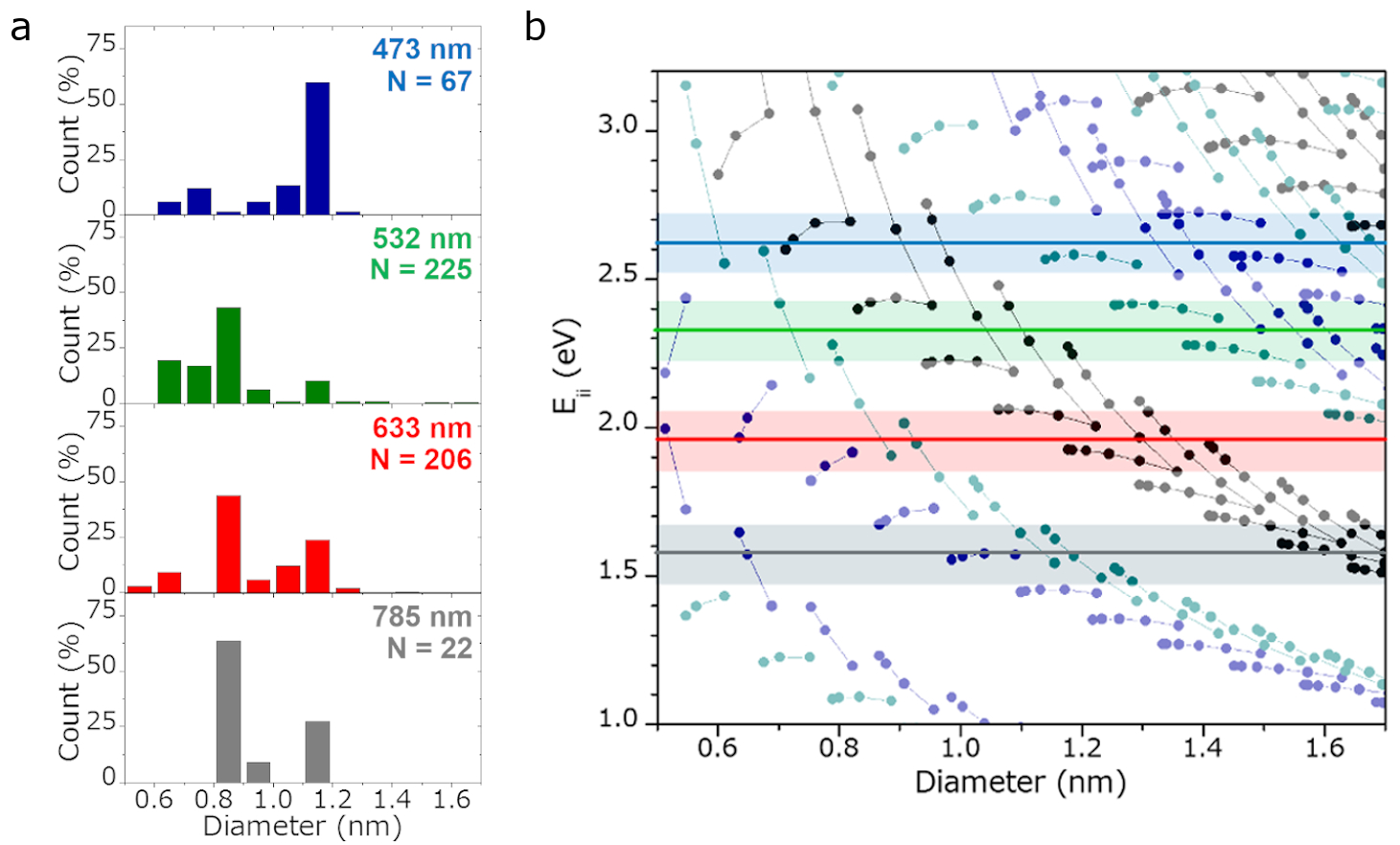}
	\caption{Confrontation of individual laser line contributions to the Kataura plot as described in Ref.~\citenum{araujo2008role}: a) Individual contributions of each laser line to the micro Raman mapping-extracted diameter distribution of Sample 1 (from top to bottom: 473 nm, 532 nm, 633 nm, 785 nm) normalized for each laser line by the total number of observations N, b) Kataura plot showing each laser line (colored lines with corresponding resonance windows delimited by the lighter rectangle around them). The resonance windows are here for visualization purposes and were not determined experimentally. The blue points correspond to E$_{ii}$ for semiconducting SWCNTs (dark blue for type I, and cyan for type II), and black points for metallic SWCNTs.}
	\label{dt-sorted-RamanvsKataura}
\end{figure*}
To explain the observed substantial differences between TEM and Raman characterizations, we looked at the contribution of each laser line to the global Raman-extracted distribution, confronted with the Kataura plot (see Figure \ref{dt-sorted-RamanvsKataura}). The Kataura plot used here is described in \cite{araujo2008role}, applying the 40 meV redshift determined in the paper to account for environmental effects, the relevance of which is discussed further in the text. Four possible influencing factors can be considered to explain the shape of the distribution: the patchy vision of the Kataura plot given by the use of only four discrete laser lines (see Figure \ref{dt-sorted-RamanvsKataura}.b), the wavelength-dependent effect of the SiO$_2$ layer thickness of the substrate on RBM intensity, the chirality and diameter-dependence of Raman cross-sections, and/or unknown chirality enrichments in the sample.\\\
It is known that interference phenomena lead to wavelength-dependent attenuation or enhancement of the Raman signal of SWCNTs sitting on a SiO$_2$/Si substrate, which depend on the thickness of SiO$_2$ used \cite{monniello2019comprehensive, yoon2009interference, wang2008interference}. For the characterization of Sample 1, a substrate with a 300 nm thick SiO$_2$ layer was chosen, as it is often used in studies on CVD growth of SWCNTs. It can be seen on the histogram in Figure \ref{dt-sorted-TEM-Raman}.a, that the distribution is dominated by the contributions from two laser lines: the 633 nm, and 532 nm lasers. As shown in Figure \ref{dt-sorted-RamanvsKataura}.a, these two laser lines allowed the detection of a total of 431 RBM peaks, representing more than 80\% of the global detected RBMs. This can be partly explained by the effect of the substrate that is more favorable for these two wavelengths compared to the 473 nm and 785 nm laser lines \cite{monniello2019comprehensive}. To investigate this effect, Raman spectra were acquired for the same SWCNT suspension (the parent HiPco sample used in the next section) dropcasted on a substrate with a 90 nm oxide layer, and one with a 300 nm oxide layer at excitation wavelengths of 476.5 nm, 632 nm and 785 nm (see Figure S4 in the ESI). The spectra show that on the 300 nm oxide substrate, there is an attenuation of the RBM intensity at  476.5 nm and 785 nm when compared to the spectra acquired on the 90 nm oxide substrate. Even though counting RBM peaks regardless of their intensity may give the impression of bypassing such issues, we assume here that this phenomenon can explain the predominance of peaks counted at wavelengths favored by interference effects.\\\
However, this cannot fully explain the shape of the distribution. Even when only looking at the two predominant contributions, some populations in the diameter range of the sample that should be detected are not seen. For instance, the bimodal shape of the 633 nm contribution could easily be explained by the crossing of two branches of the Kataura plot with the laser line. But the absence of detected nanotubes for diameters above 1.2 nm at this excitation energy does not make sense within the scheme of this simple interpretation. The case of the 532 nm laser line is also revealing: very little SWCNTs with diameters above 0.9 nm are detected, when they in fact should be detected if present. We have also verified the influence of different empirical relations of RBM-frequency versus SWCNT diameter (even those not adapted to our type of samples) on the resulting diameter distributions and found that the observed discrepancies between the two characterization techniques are not due to the choice of the empirical law (see Figure S3 in the ESI). \\\
These observations show that other factors, like chirality-dependent cross-sections and the partial vision of the Kataura plot, may have a strong influence on the results. However, it is difficult to conclude on the extent of this influence without knowing the chirality distribution in the sample, when chirality enrichments could also play a role. In order to ensure that this is not the case, a HRTEM characterization of the (n,m) distribution in the sample could be done. But the wide diameter distribution of the present sample, with over 100 possible chiralities in the 0.6 - 1.6 nm diameter range, implies that a characterization of an extremely large number of SWCNTs would be necessary in order to obtain representative statistical data.\\\
For this reason, we conducted this same comparison on two chirality-enriched samples. The narrowness of the diameter and chirality distributions for these samples allows for a representative HRTEM characterization of the composition of the sample, leading to a more thorough understanding of the potential discrepancies between TEM- and Raman peak counting-extracted diameter distributions. The Raman characterization for these samples was performed on SiO$_2$/Si substrates with the more optimal 90 nm thick oxide layer to reduce the influence of the substrate on the resulting distributions, as evidenced by Figure  S4 in the ESI.

\subsection{Chirality-enriched SWCNT samples}

\begin{figure*}[h!]
	\centering
	\includegraphics[scale=0.14]{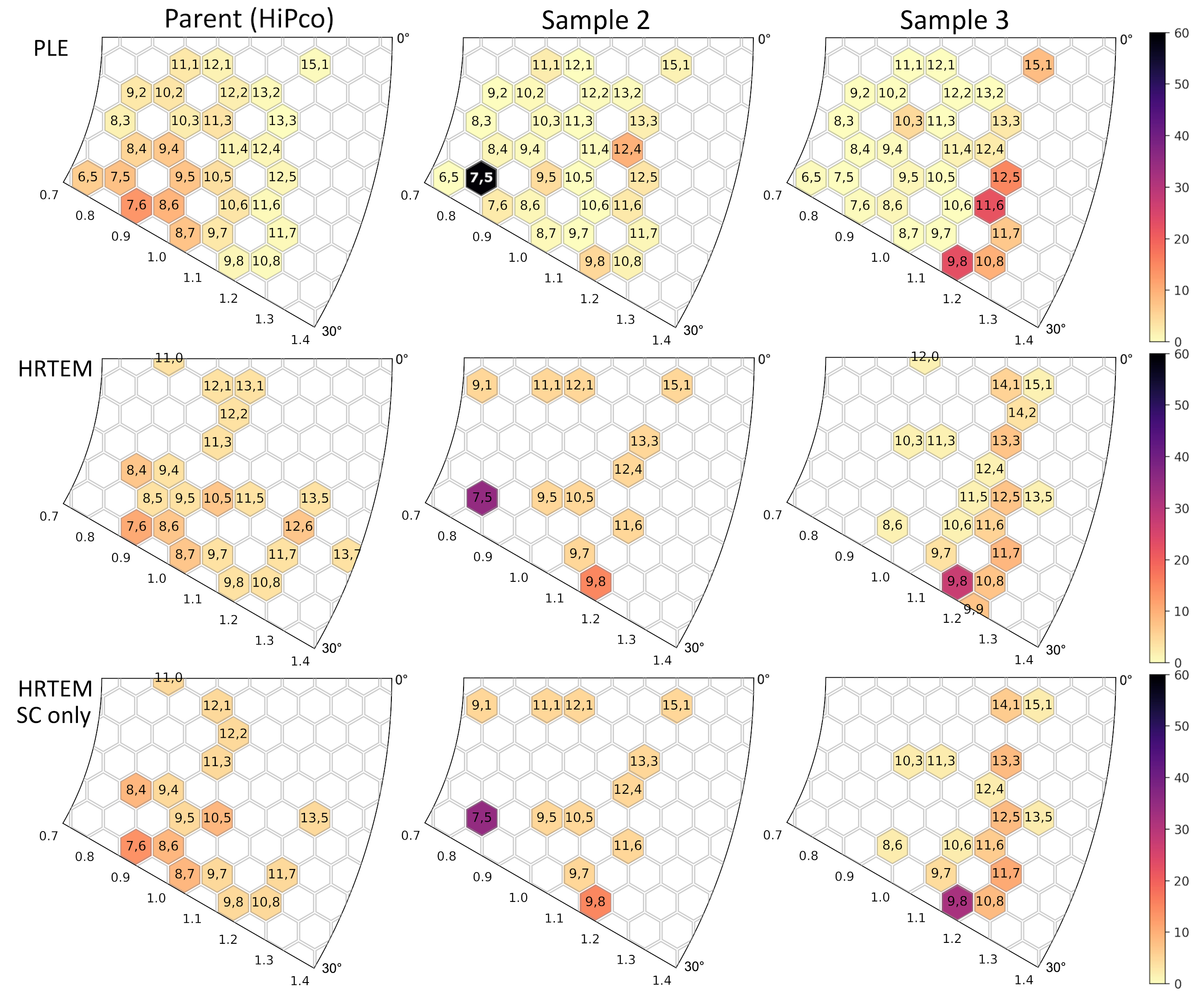}
	\caption{Composition of the two chirality-enriched samples, and the parent HiPco sample: chirality distributions obtained from (top) PLE intensities, (middle) HRTEM analysis for (from left to right) the parent HiPco sample, Sample 2, and Sample 3. The bottom panels give the chirality distribution obtained from HRTEM without the metallic SWCNTs. The relative abundances and PL intensities of each chirality are represented in percentages, on a diameter-chiral angle polar plot. The HRTEM distribution without metallic tubes can be directly compared with the PLE intensity distribution.}
	\label{nm-sorted-PLE-HRTEM}
\end{figure*}

\begin{figure*}[h!]
	\centering
	\includegraphics[scale=0.125]{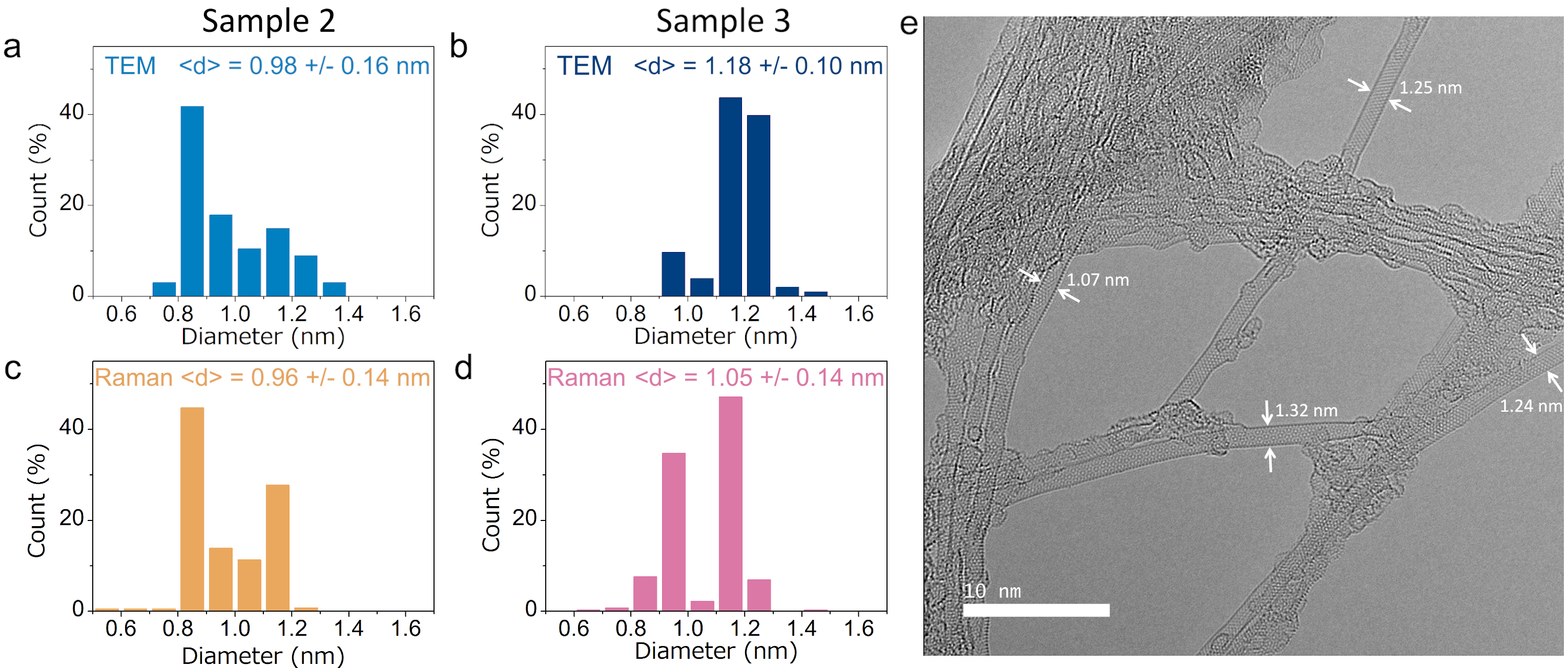}
	\caption{Comparison of TEM- and micro Raman mapping-extracted distributions for the chirality-enriched samples: a) and b) TEM-extracted diameter distribution histograms for Sample 2, and Sample 3, respectively. c) and d) Raman-extracted diameter distribution histograms for Sample 2, and Sample 3, respectively. The mean diameter and standard deviation of the width of the distribution are given for each distribution. e) Example of a HRTEM image taken from Sample 3, exemplifying the narrowness of the diameter distribution.}
	\label{nm-sorted-RamanvsTEM}
\end{figure*}

Here, we used a (7,5)-enriched sample, and a sample with a narrow chirality distribution with diameters around 1.2 nm (centered around the (9,8) and (11,6) chiralities) obtained after aqueous two-phase sorting of the same HiPco SWCNTs parent suspension, according to the procedure detailed in Ref.~\citenum{subbaiyan2014role} and Ref.~\citenum{VanBezouw2018}, respectively (see Methods section for experimental details and Figure S6 for UV-VIS spectra). These samples will further be referred to as Sample 2 and Sample 3, respectively. The compositions of the samples were also determined by wavelength-dependent near-IR photoluminescence-excitation spectroscopy (PLE) and by a statistical HRTEM study (see Figure S5 in the ESI for procedure). In PLE, a technique which is also very frequently used for chirality assessment of SWCNTs, the emission of semiconducting SWCNTs is mapped out versus the higher optical transitions of the SWCNTs. Thereby the observed intensities are both influenced by the relative absorption cross-section of the SWCNTs (excitation) and their PL quantum efficiency (emission) and hence chirality distributions extracted from PLE are modulated by these two effects. To extract these intensity distributions, we fitted the experimental PLE maps of the two samples and the parent HiPco sample simultaneously with the same peak positions and linewidths, but with varying amplitudes using the previously developed empirical fitting model as in Ref.~\citenum{cambre2015asymmetric, li2019separation} (see Figure S7 and S8 in the ESI for PLE experimental and fitted maps, and details of fitted amplitudes, respectively). The PLE map of the parent HiPco sample was fitted along with those of the two sorted samples so that all chiralities present in the initial sample (\textit{i.e.} the chiralities that can potentially end up in the sorted samples), can be fitted and their intensity can be extracted, even if they are no longer present in the sorted samples. In the absence of sufficient data regarding chirality-dependent PL cross-sections, the abundance of a chirality was directly correlated to its PL intensity. The quantitative results are shown, along with the results for the exact same samples extracted from HRTEM analysis, in Figure \ref{nm-sorted-PLE-HRTEM}. Using HRTEM takes into account the presence of metallic SWCNTs in all samples, which are overlooked when only using a PLE analysis.\\\
Though the analysis of the HRTEM statistical data is done on only tens of nanotubes (see Table S1 in the ESI for detailed results) and cannot be fully reliable for determining the exact quantitative composition of the sample, it allows for confirming the trends observed in the PLE results. Globally, the two techniques gave consistent results: Sample 2 is well chirality-sorted, with a clear predominance of the (7,5) chirality, with several other chiralities present within the diameter range of the parent HiPco sample (0.7 - 1.3 nm), and Sample 3 is mostly composed of various chiralities within a narrow diameter range (1.1 - 1.3 nm, with the (9,8) and (11,6) SWCNTs among the most abundant chiralities). The (7,5) chirality represents almost 60\% of the intensity of the detected SWCNTs using PLE, and only 35\% using HRTEM in Sample 2. This can be attributed to the fact that the intrinsic PL intensity of the (7,5) SWCNT is theoretically several times higher than that of the (9,8) or the (11,6) for instance \cite{oyama2006photoluminescence}. The main difference between the two techniques, aside from the detection of metallic SWCNTs in HRTEM, is that the smallest-diameter SWCNTs (namely the (6,5) and (8,3)) which are more challenging to observe in HRTEM are not reported in the HRTEM results, and HRTEM allows the detection of larger-diameter SWCNTs that are not seen in PLE (like the (13,5)). \\\
First, the TEM- and Raman-extracted diameter distributions were compared for Sample 2. The two diameter distribution histograms are presented in Figure \ref{nm-sorted-RamanvsTEM}.a,c (see Figure S9 in ESI for Raman spectra). With a 2\% relative difference between the mean diameters, the two distributions seem very close. Additionally, the shapes of the distributions are quite similar: both show a majority population of SWCNTs in the 0.8-0.9 nm diameter range, which is in good agreement with the abundance of the (7,5) chirality in Sample 2 (0.82 nm diameter, taking a 1.42 \AA{} carbon-carbon distance). The bimodal aspect of the Raman-extracted distribution is sharper than for the TEM-extracted distribution, with a slight overestimation of the SWCNT populations in the 1.1-1.2 nm range, but contrary to the case of Sample 1, there are no major discrepancies. The fairly good agreement between the two characterization methods for this specific sample shows that the Raman based methodology can actually lead to an accurate assessment of the diameter distribution, when selecting a correct set of laser lines for each particular sample. However, the choice of laser wavelengths is impossible to make without pre-knowledge on the sample composition.\\\
Then, we compared the two distributions obtained for Sample 3. According to TEM (Figure \ref{nm-sorted-RamanvsTEM}.b), the diameter distribution of the sample is in very good agreement with the PLE and HRTEM-extracted chirality distributions: the diameter distribution is extremely narrow, with a standard deviation of 0.10 nm, and centered around a mean value of 1.18 nm, very close to the (9,8) and (11,6) diameters of 1.15 nm and 1.17 nm, respectively. The very efficient diameter selectivity of the sorting for this sample is well shown on the HRTEM image given in Figure \ref{nm-sorted-RamanvsTEM}.e.\\\
When looking at the Raman-extracted distribution for this sample (see Figure \ref{nm-sorted-RamanvsTEM}.d, and Figure S10 in ESI for Raman spectra), we can notice the appearance of a predominant population of SWCNTs within the 0.9-1.0 nm diameter range, which is almost absent in the TEM-extracted distribution, leading to a significant shift of the mean diameter towards lower values. Moreover, almost half of the SWCNTs observed by TEM in the 1.1-1.3 nm range do not seem to be accounted for by our Raman methodology. The case of this chirality-enriched sample is similar to Sample 1: an overall agreement on diameter range, but the bimodal shape of the Raman-extracted distribution can lead to substantial errors in the estimated distribution. Here, a chirality distribution based on the Raman data would lead to tremendously misleading results.\\\
\begin{figure*}[h!]
	\centering
	\includegraphics[scale=0.28]{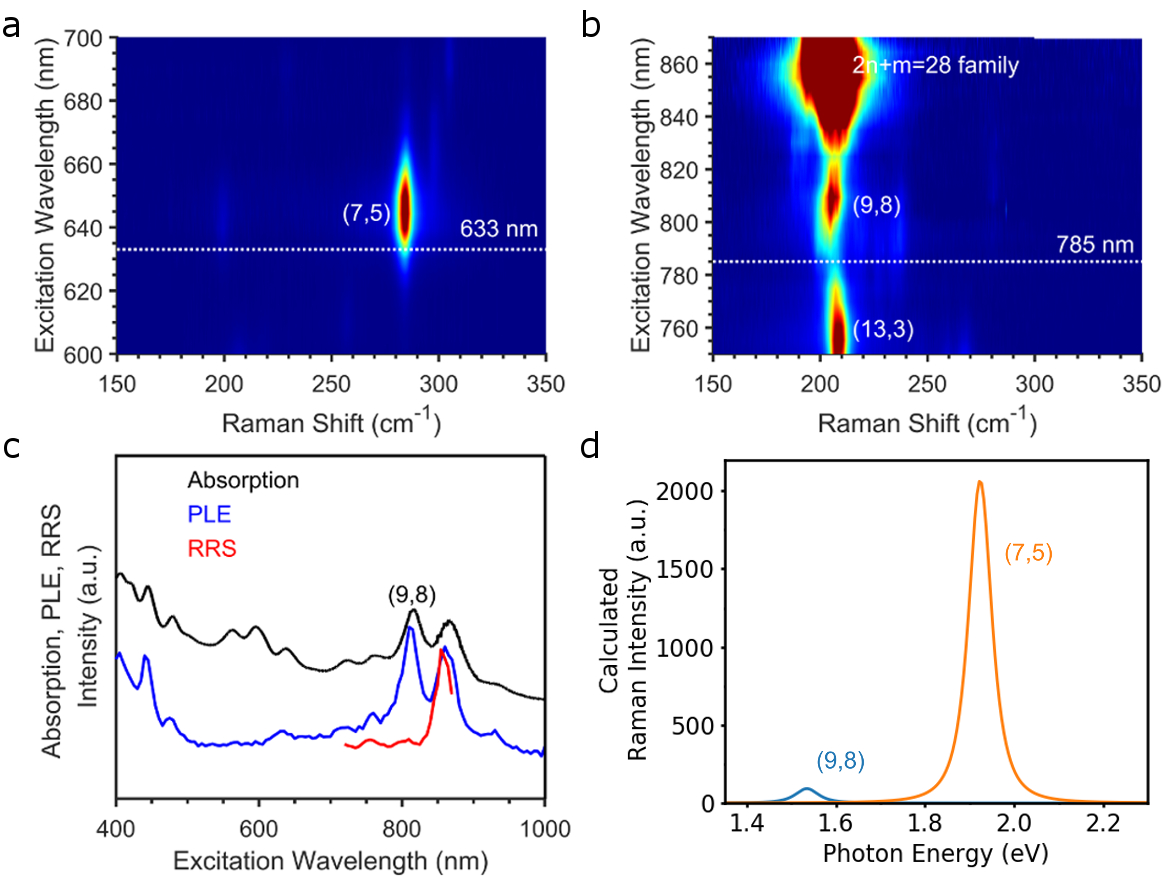}
	\caption{Resonance conditions of the (7,5) and (9,8) chiralities: Wavelength-dependent Raman maps with 5 nm stepsize obtained in liquid suspension for a) Sample 2 zoomed-in on the region where the E$_{22}$ of the (7,5) SWCNT is resonant, and b) for Sample 3 zoomed-in on the region where the E$_{22}$ of the (9,8) SWCNT is resonant. c) Optical absorption (black), Normalized integrated PL-excitation spectrum obtained by integrating the PLE maps between emission wavelengths 1100-1700 nm (blue) and normalized integrated RRS-excitation spectra obtained by integrating the Raman maps between 150-350 cm$^{-1}$ (red), d) calculated Raman resonance profiles for the (9,8) and (7,5) SWCNTs according to Ref.~\citenum{popov2004resonant}.}
	\label{98-75-Raman-cross-sections}
\end{figure*}
The Raman data were analyzed further to better understand the observed discrepancies, based on the knowledge of the chiralities present in the samples from PLE and HRTEM, and the use of an adapted Kataura plot. First, to ensure that the Kataura plot used was sufficiently adapted to the samples, wavelength-dependent Raman measurements were performed on Sample 2 and Sample 3 in solution, around the E$_{22}$ of the (7,5), and (9,8) chiralities, respectively (see Figure \ref{98-75-Raman-cross-sections}.a,b). For (7,5) and (9,8) chiralities, the maximum Raman intensities were seen at 644 nm and 811 nm, respectively. This corresponds to E$_{22}$ values of 1.925 eV for the (7,5), and 1.529 eV for the (9,8), which are close to the calculated values of 1.915 eV (10 meV redshift), and 1.544 eV (15 meV blueshift). The resonance profiles of the (7,5) nanotubes from Sample 2 were then compared in solution and deposited on the SiO$_2$/Si surface (see Figure S11 in the ESI) to ensure that the same Kataura plot can be used to interpret our RBM peak counting results (proving that SWCNTs are still surrounded by surfactant in the deposited sample). Moreover, the RBM frequencies did not appear to be significantly shifted when going from a dispersed sample to the deposited one (see Raman spectra comparison in Figure S11 in the ESI), ensuring that the RBM-to-diameter law used is correct. Proof of the absence of doping from the substrate is also provided in Figure S12 in the ESI, which shows that there is no shift between the G-band frequencies for the parent HiPco sample measured in solution and on a SiO$_2$/Si substrate (90 nm oxide) \cite{tsang2007doping}.\\\
\begin{figure*}[h!]
	\centering
	\includegraphics[scale=0.19]{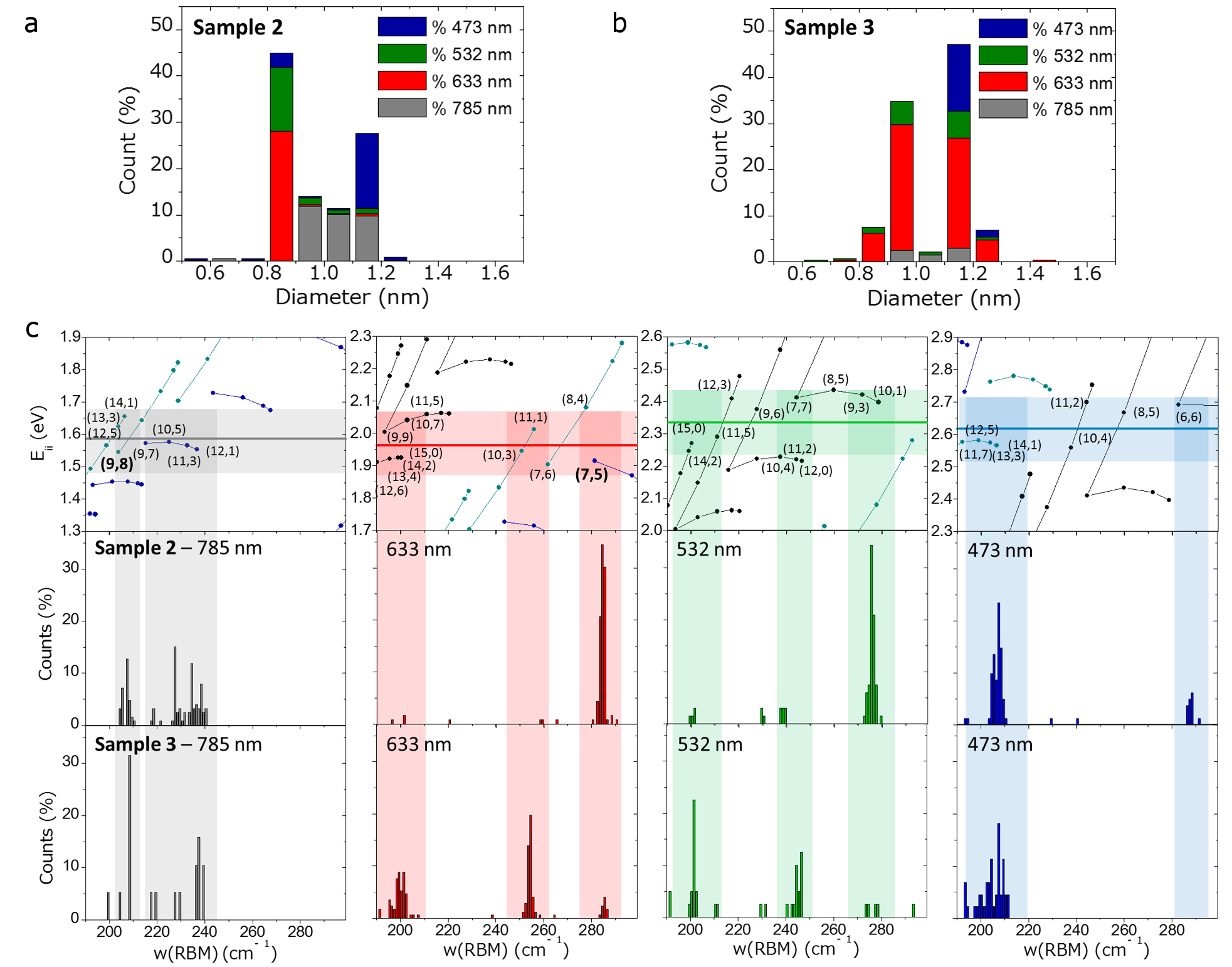}
	\caption{Comparison of individual laser line contributions to diameter distributions with the Kataura plot: a), and b) Raman peak counting-extracted diameter distribution of Sample 2, and Sample 3, respectively, showing the contributions of each laser line to each bin. c) From top to bottom : corresponding section of the Kataura plot, RBM frequency counts for Sample 2, and RBM frequency counts for Sample 3, normalized for each laser line. On each section of the Kataura plot, the energy of the laser line is shown by the colored line in the center, and the lighter rectangle gives a wide possible resonance window. The (n,m) for all chiralities falling in this window for each laser line are given for clarity.}
	\label{nm-sorted-RamanvsKataura}
\end{figure*}
On this basis, details of the RBM peak counting-extracted diameter distributions shown in Figure \ref{nm-sorted-RamanvsKataura} are analysed as follows. Figure \ref{nm-sorted-RamanvsKataura}.a,b show the diameter distributions normalized by laser line contributions for Samples 2 and 3, respectively, and Figure \ref{nm-sorted-RamanvsKataura}.c gives the corresponding RBM frequency distributions for each laser line, and each sample, confronted with the corresponding portion of the Kataura plot. Looking at Sample 2 (Figure \ref{nm-sorted-RamanvsKataura}.a), the peak at 0.8 - 0.9 nm can be mainly attributed (by looking at Figure \ref{nm-sorted-RamanvsKataura}.c) to the resonance of the (7,5) chirality at 633 nm (at about 285 cm$^{-1}$), and of the (9,3) and/or (10,1) at 532 nm (around 278 cm$^{-1}$). Since the (7,5) is predominant in the sample based on HRTEM and PLE, its strong presence in the Raman distribution is consistent. This is due to its fairly high Raman cross-section and a good choice of the excitation wavelength (see Figure \ref{98-75-Raman-cross-sections}.d for the calculated intensity profile according to Ref.~\citenum{popov2004resonant}, and Figure \ref{98-75-Raman-cross-sections}.a for proof of the resonance conditions of the (7,5) chirality at 633 nm). The presence of the (9,3) and/or (10,1) here, and their absence in the HRTEM data can be explained by the low number of SWCNTs studied with HRTEM and the higher damage of the electron beam for such small diameter SWCNTs, while their substantial contribution to the distribution could also stem from their expected high Raman cross-sections (lower branch transitions, and near zigzag chiralities \cite{sato2010excitonic}). Between 0.9 - 1.2 nm, we can assume a contribution of the $2n + m = 29$ (around 205 cm$^{-1}$), and the $2n + m = 25$ families (between 215 - 245 cm$^{-1}$). The corresponding chiralities are overall detected by the HRTEM and PLE results combined (see Figure S8 and Table S1 in the ESI). Interestingly, the slight overestimation of the SWCNT populations in the 1.1 - 1.2 nm diameter range can easily be attributed to the fact that the SWCNTs in the $2n + m = 29$ family are actually probed twice with our setup ((14,1), (13,3), and (12,5) nanotubes), at 785 nm (E$_{22}$), and at 473 nm (E$_{33}$). Taking this into account by removing the contribution of the 473 nm laser line would result in a more accurate diameter distribution.\\\
In the case of Sample 3, several discrepancies have to be studied, but we can also point out some consistencies. First, looking at the contribution from the 633 nm laser line, it can be seen that the (7,5) chirality is much less detected than for Sample 2, which brings down the peak in the distribution at 0.8 - 0.9 nm, consistent with the HRTEM- and PLE-extracted compositions, as well as the TEM-extracted diameter distribution. Then, the predominant peak in the distribution at 1.1 - 1.2 nm, \textit{i.e.} corresponding to an RBM frequency of approximately 200 cm$^{-1}$, is mainly attributed to the resonance of the (9,9) and (14,2) nanotubes (which together represent 11\% of the HRTEM chirality distribution with a total of six SWCNTs counted) at 633 nm. Here again, the contributions from the 532 nm and 473 nm laser lines are attributed to chiralities that can already be detected at 633 nm, and 785 nm, respectively, meaning that the SWCNT population in this diameter range might also be slightly overestimated.\\\
As for the inconsistencies observed in Figure \ref{nm-sorted-RamanvsTEM} in comparison with TEM, we will focus on the two main ones: the substantial increase in the population between 0.9 - 1.0 nm, and the drastic decrease in the population between 1.2 - 1.3 nm. It is striking to see that the SWCNTs detected in the 0.9 - 1.0 nm diameter range almost entirely stem from the contribution of the 633 nm laser line, which could be due to the presence of two chiralities: the (10,3) and (11,1). Looking at the HRTEM and PLE data in Figure \ref{nm-sorted-PLE-HRTEM} (and Figure S8 and Table S1 in the ESI), we can deduce that the (10,3) and (11,1) chiralities are present in the sample. Though it is only barely detected in PLE, the presence of the (11,1) chirality in the sample cannot be ruled out. \\\
To examine this assumption, wavelength-dependent Raman measurements were performed in solution for Sample 3 between 600 nm and 700 nm in the RBM spectral range (see Figure S13 in the ESI). We can clearly see that the RBM peaks detected at 633 nm between 250 and 260 cm$^{-1}$ can be attributed to the (10,3) chirality. Therefore, this major discrepancy is due to the resonance of one SWCNT chirality, which represents less than 5\% of the global population of the sample according to HRTEM and PLE (see Figure S8 and Table S1 in the ESI). One can note that this peak does not appear in the case of Sample 2, even though it has a higher content of (11,1) nanotubes than Sample 3 according to PLE and HRTEM, which is consistent with the fact that what we observe in Sample 3 is due to the resonance of (10,3) nanotubes. Here, the joint effects of the fairly high Raman cross-section \cite{popov2004resonant}, and the closeness of the E$_{22}$ optical transition energy of the (10,3) SWCNT to the excitation energy leads to the predominance of the 0.9 - 1.0 nm population. This could have led, in the case of a chirality distribution assessment based on Raman peak counting, to the nearly tenfold overestimation of the (10,3) population. \\\
To explain the second inconsistency, we have to look at the chiralities present in the sample in this diameter range, that are not detected within our discrete wavelength Raman setup. According to our HRTEM and PLE characterizations, the following chiralities with diameters between 1.2 nm and 1.3 nm are present in the sample: (15,1), (9,9), (10,8), (11,7) and (13,5), representing about 30\% of all SWCNTs (see Figure S8 and Table S1 in the ESI). According to the Kataura plot, only two out of five of these chiralities are detectable with our discrete wavelength Raman setup. The (15,1), (10,8) and (13,5) SWCNTs all have optical transition energies that fall outside of the resonance windows for all the laser lines used in this study. The (9,9), and (11,7) SWCNTs appear to be detected by the 633 nm and the 473 nm lasers, and their presence may be slightly underestimated because of their low cross-sections \cite{sato2010excitonic}. \\\
Finally, it is also interesting that the (9,8) chirality, which is one of the predominant chiralities in this sample according to HRTEM and PLE, is not detected. Contrary to the case of Sample 2, where the predominant chirality "leads" the diameter distribution in PLE, HRTEM and Raman imaging, here 1.1 - 1.2 nm population is predominant but attributed to other chiralities. This lack of detection of the (9,8) nanotube can be interpreted by looking back at Figure \ref{98-75-Raman-cross-sections}. According to the wavelength-dependent Raman map shown in Figure \ref{98-75-Raman-cross-sections}.b, the 785 nm laser line falls in the tail of the (9,8) resonance profile (50 meV distance between its E$_{22}$ and the excitation energy). However, this is also the case for the (7,5) chirality, which is well detected in Sample 2 even if the excitation wavelength does not fall at the highest intensity in the (7,5) resonance profile (33 meV distance). It seems that what makes the (9,8) undetectable is the coupled effects of non-ideal resonance conditions and of its low Raman cross-section (see Figure \ref{98-75-Raman-cross-sections}.d for comparison with the (7,5) SWCNT). This is also confirmed experimentally: Figure \ref{98-75-Raman-cross-sections}.c shows the comparison of optical absorption spectroscopy, PLE, and RRS spectra acquired from Sample 3. While the absorption and PL-excitation profiles show a similar ratio of the (9,8) peak (\textit{i.e.} 810 nm) with respect to the peak of the 28-family (\textit{i.e.} observed at 850nm) the RRS-excitation profile hardly shows a resonance for the (9,8) chirality, evidencing the very low theoretically predicted Raman cross-section of the (9,8) SWCNT \cite{popov2004resonant}.\\\
Overall, it appears that the discrepancies observed for Sample 3 can be explained by the patchy vision of the Kataura plot given by only four excitation wavelengths, and chirality-specific Raman cross-sections. With this in mind, it also seems that the fairly good agreement between TEM and Raman imaging for Sample 2 is almost accidental.

\subsection{Elements of improvement for RBM peak counting}
The idea behind the RBM peak counting methodology, aside from experimental convenience and speed, is to bypass the chirality-dependence of Raman cross-sections and the influence of the chosen substrate on RBM intensity as a function of excitation wavelength. Since the intensity of peaks is not taken into account, it is assumed that SWCNTs with lower cross-sections, or detected in unfavorable optical conditions will be counted even if their RBM peaks have a low intensity. This study shows that this assumption is misleading because it appears that these effects have a strong influence on the RBM peak counting diameter distributions.\\\ 
For an improved routine use of the RBM peak counting methodology, we propose the following procedure. First, select the adapted substrate with an oxide thickness that will not induce drastic wavelength-dependent RBM intensity attenuation or enhancement depending on the available excitation wavelengths. Then, an appropriate set of laser wavelengths for Raman imaging has to be selected. Such a choice, with optimal coverage of all chiralities, while minimizing double counting of certain chiralities, can only be made by taking into account 3 factors: (1) the positions of the electronic transitions, E$_{ii}$ of the SWCNTs, or in other words how to rescale the Kataura plot properly for the sample under investigation; (2) the widths of these electronic resonances and (3) combine this with the theoretically predicted Raman cross-sections (see for instance empirical formulas in reference \citenum{popov2006resonant}: if cross-sections of some chiralities are very low at one of the chosen wavelengths, then either exposure times need to be increased to the point where detection of these chiralities is nevertheless ensured, or another wavelength covering a different transition of those chiralities needs to be included). The first two factors are mainly determined by the (external and internal) environment of the SWCNTs and are ideally obtained through wavelength-dependent Raman spectroscopy over a wide wavelength range. However, as this technique is not widely available, even probing a narrow range around one of the expected resonances with a tunable laser combined with a few additional discrete laser wavelengths can be sufficient to rescale the Kataura plot for a representative sample in the given environment, for example SWCNTs synthesized on a substrate or in a vertical forest, isolated SWCNTs or SWCNT bundles, etc. In fact, many of such rescaled Kataura plots can already be found in the literature \cite{araujo2008role, liu2012atlas, weisman2003dependence}. If the diameter range of a sample is known beforehand (e.g. through other characterization techniques such as the proposed TEM characterization in this work or absorption spectroscopy and/or fluorescence-excitation spectroscopy), the number of required Raman wavelengths may be further reduced, simplifying the Raman peak counting methodology.

\section{Conclusion}
In conclusion, we have demonstrated the extreme care that should be taken when interpreting selectivity assessment results based solely on RBM peak counting from micro-Raman mapping on flat substrates. Using a SWCNT sample with a broad chirality distribution (Sample 1), we first showed discrepancies between TEM and Raman characterizations when considering the diameter distribution of the sample. With a more in-depth study on two chirality-enriched samples, we were able to show the strong joint influence of chirality-dependent Raman cross-sections, and resonance conditions in the resulting Raman peak counting-extracted diameter distributions. Though in one case, where the predominant chirality was easily detectable by Raman (high cross-sections and the right choice of excitation wavelengths), the Raman-extracted diameter distribution seemed consistent with TEM experiments (Sample 2), substantial discrepancies were seen when this was not the case (Sample 3). The observed discrepancies were further investigated by comparing the Raman- and TEM-extracted diameter distributions with chirality distributions obtained by PLE and statistical HRTEM experiments. This work is, to our knowledge, the first to report a comparison between RBM peak counting with four excitation wavelengths and TEM for diameter distributions of samples, and the first to report a cross-correlated chirality distribution assessment from PLE and statistical analysis of HRTEM images. In addition, the Raman intensities from wavelength-dependent RRS on the solution samples were compared to absorption and PL-excitation intensities, showing large variations in Raman cross-sections depending on the specific chirality studied. Our results also showed that the smallest SWCNTs in the samples were more challenging to detect in TEM. This work clearly shows the crucial importance of cross-characterization to avoid any misleading interpretations when making purity claims, and the need to continue investigating characterization techniques for SWCNT samples, with the goal to develop standardized metrology procedures.

\section{Experimental}
\label{Exp}
\subsection{Sorting methods}
Sample 1 was prepared from CNTs that were grown using CVD with methane (CH$_4$) as a carbon source, and a Co:Mo-MgO catalyst \cite{flahaut2000synthesis}, which leads to the growth of 77$\%$ DWCNTs, giving access through sorting, to broad-diameter distribution SWCNT samples. The as-produced nanotubes were dispersed by tip sonication (1 mg/mL) in an aqueous solution containing 2 wt$\%$ of sodium cholate (SC) and were centrifuged for one hour at 35000 rpm. The extracted supernatant was sorted using the DGU technique described in \cite{fleurier2009sorting}. Sample 1 corresponds to the first layer of the DGU which contains more than 96$\%$ of SWCNTs, with a broad diameter distribution. \\\
For Sample 2 and Sample 3, the parent HiPco dispersion was prepared by solubilizing the SWCNTs (NoPo Nanotechnologies (batch number-2015.820), Inc.) in a 1\% wt sodium deoxycholate (DOC, Acros Organics, 99\%) solution in D$_{2}$O (Cortecnet, 99.8 atom \% D) (10 mg/ml). The samples were stirred for 3 weeks, and were bath sonicated for 15 minutes on the first 3 days of stirring. Afterwards, the samples were centrifuged for 4 hours at 16000g (14000rpm, Sigma 2-16KCH centrifuge, with swing-out rotor) to remove all non-dissolved species, yielding the parent HiPco sample. (7,5) and (9,8) ATPE enriched samples, were separated by acqueous two-phase sorting of the parent HiPco dispersion, using the previously described two-step sorting procedures for (7,5) \cite{subbaiyan2014role} and (9,8) SWCNTs \cite{VanBezouw2018}. Afterwards, the samples were concentrated and dialysed at least 10 times using centrifuge filters with a 100kD pore membrane (Amicon) to remove the polymers and surfactant mixture used in the ATPE separation and end up with SWCNTs in a 1\% DOC/D$_2$O solution.

\subsection{Transmission electron microscopy}
The TEM grids for all samples were prepared as follows: several drops of the SWCNT dispersion were dropped onto a copper TEM grid covered by a holey carbon film, and left to dry in ambient air for two hours. The grid was then rinsed with large amounts of deionized (DI) water to remove excess surfactants on a Buchner setup mounted with a glass column, and dried in ambient air.\\\
In the case of Sample 1, all TEM observations were made on a Zeiss Libra microscope equipped with a monochromated electron gun and an energy filter to reduce chromatic aberrations. The microscope was operated at 80kV in order to reduce knock-on damage. For the diameter-enriched samples, atomically resolved HRTEM was performed on an aberration-corrected Jeol JEM-ARM-200F microscope also operating at 80 kV. The low energy dispersion of the cold FEG source associated with the image corrector delivers a spatial resolution below 150 pm and allows the observation of the honeycomb lattice. The method used for chirality assessment can be found in the supplementary information (Figure S5). Simulated HRTEM images used for (n,m) assignment were calculated within dynamical theory with the commercial code JEMS \cite{stadelmann1987ems}, and electron diffraction patterns were calculated using the open-source code DiffractX \cite{kirkland2010advanced, kociak2003accurate}.\\\
To determine the diameter distribution for each sample, we measured the diameters for each clean SWCNT suspended over vacuum on every TEM image. Diameter measurements were extracted from an intensity profile perpendicular to the tube axis as described in \citenum{fleurier2009transmission}.
 
\subsection{Raman spectroscopy}

\subsubsection{RBM peak counting}
The samples for the Raman peak counting method were prepared on SiO$_2$/Si wafers (300 nm thermal SiO$_2$ layer for Sample 1, 90 nm thermal SiO$_2$ layer for Samples 2 and 3). Wafers were sonicated successively in acetone, isopropanol, and dichloromethane for 5 minutes, then underwent a O$_2$ plasma treatment for 10 minutes. A drop of the SWCNT dispersion was added on the surface of the wafer, and spread using a N$_2$ flow, then placed on a hot plate at 50$^\circ$C for 5 minutes to dry. The surface was thoroughly rinsed under a stream of DI water to remove excess surfactant, and placed back on the hot plate for 5 minutes. \\\
For the "RBM counting" experiments, all Raman spectroscopy characterizations were performed using a HORIBA LabRam ARAMIS spectrometer, using a x100 objective, with four excitation wavelengths (473 nm, 532 nm, 633 nm, and 785 nm). The spectra were acquired for 1s using a 1800 groove per mm grating. \\\
In order to obtain reliable statistical data for diameter distribution assessment, we proceeded systematically with the same method. Raman spectrum grids were recorded on the surface of the characterized wafer. The dimensions of the grids were 50 $\mu$m by 50 $\mu$m, with a 5 $\mu$m step in both directions, which amounted to a total of 121 recorded spectra for each grid. We used a large enough step to avoid detecting the same SWCNT twice in a row, but small enough to avoid going over a too large area in one grid, and possibly losing focus conditions. The grids were done on random locations on the substrate with the four excitation wavelengths. The same number of spectra were recorded with each laser. The laser power and acquisition time were adjusted in order to have significant RBM signal in a short enough time without heating the sample. The spectrometer was calibrated for a SiO$_2$ signal at 520.7 cm$^{-1}$. Each RBM peak with an intensity larger than three times the background noise was taken into consideration and treated as representative of one resonating SWCNT. The diameter of the corresponding SWCNT was determined using the following relation:
\begin{equation}
\omega_{RBM} = \frac{227}{d_t}\sqrt{1 + C_ed_t^2}
\label{eqn:RBM2007}
\end{equation}
Where $\omega_{RBM}$ is the RBM frequency, $d_t$ the diameter, and $C_e$ a constant representing the environmental effects. Here, a $C_e$ value of 0.05786 nm$^{-2}$ was used \cite{araujo2008nature}. 
\\\ 
\subsubsection{Wavelength-dependent Raman spectroscopy}
To compare the RBM-frequencies and excitation profiles of (7,5) and (9,8) samples in solution and as deposited on the substrate, resonant Raman spectra were collected in backscattering geometry using a Dilor XY800 triple-grating spectrometer equipped with a liquid-nitrogen cooled CCD detector. Excitation wavelengths were obtained from a tunable Ti:sapphire (700-900nm) laser and dye laser equipped with DCM laser dye (630-690nm), both pumped by an Ar$^+$ ion laser. Individual spectra were corrected for laser intensity, detector and spectrograph sensitivity and obtained with sub-wavenumber spectral resolution.

\subsection{PLE}
2D PLE maps were collected in an in-house developed dedicated spectrometer consisting of a pulsed Xe-lamp (Edinburgh Instruments, Xe900-XP920) for excitation and a liquid-nitrogen cooled extended InGaAs array detector (Princeton Instruments OMA V:1024) sensitive up to 2.2 $\mu$m. Spectra were recorded in 90$^{\circ}$ geometry in a 3 mm microcell, with 5 nm steps in excitation wavelength. Appropriate filters were used to eliminate stray light and higher order diffractions from the spectrometers and all spectra were corrected for detector and spectrograph efficiency, filter transmission, re-absorption within the cell and (temporal and spectral) variations of the excitation light intensity.

\section*{Acknowledgements}
The research leading to these results has received funding from the French Research Funding Agency under grant no. ANR-13-BS10-0015 (SYNAPSE project) and under grant no ANR-18-CE09-0014-04 (GIANT project), the European Union Seventh Framework Programme (FP7/ 2007e2013) under grant agreement 604472 (IRENA project). The authors also acknowledge funding from the Fund for Scientific Research Flanders (FWO projects No. 1512716N, G040011N, G021112N, and G036618N. D.L. acknowledges the FWO Postdoc Fellowship No. 12ZP720N). S.F., J.D., D.L. and S.C. also acknowledge funding from the European Research Council Starting Grant No. 679841 (ORDERin1D). The authors thank the METSA research foundation for giving access to the Cs-corrected TEM of the MPQ-Paris Diderot laboratory, the LPICM and IPVF for giving access to their Raman spectrometer. The authors acknowledge M. Kociak (DiffractX) and Y. Le Bouar for use of their software, G. Wang for assistance with Cs-corrected TEM use, C. Moreira Da Silva and L. Orcin-Chaix for technical assistance. J. S. Lauret and V. Jourdain are acknowledged for fruitful discussions.

%% The Appendices part is started with the command \appendix;
%% appendix sections are then done as normal sections
%% \appendix

%% \section{}
%% \label{}

%% If you have bibdatabase file and want bibtex to generate the
%% bibitems, please use
\bibliographystyle{elsarticle-num} 
\bibliography{Article-AC-TEM-Raman-sf}

\end{document}

\begin{frontmatter}

\title{Supplementary Information: Assessing the reliability of the Raman peak counting method for the characterization of SWCNT diameter distributions: a cross-characterization with TEM}

\author[LEM]{Alice Castan\corref{cor1}}
\cortext[cor1]{Corresponding authors. \\E-mail: alicecastan1@gmail.com, sofie.cambre@uantwerpen.be, annick.loiseau@onera.fr}
\author[ECMPL]{Salom\'e Forel}
\author[LEM]{Fr\'ed\'eric Fossard}
\author[ECMPL]{Joeri Defillet}
\author[LEM]{Ahmed Ghedjatti}
\author[ECMPL]{Dmitry Levshov}
\author[ECMPL]{Wim Wenseleers}
\author[ECMPL]{Sofie Cambr\'e\corref{cor1}}
\author[LEM]{Annick Loiseau\corref{cor1}}
\address[LEM]{Laboratoire d'Etude des Microstructures, CNRS-ONERA, University Paris-Saclay, Ch\^{a}tillon, France}
\address[ECMPL]{Experimental Condensed Matter Physics Laboratory, University of Antwerp, Belgium}

\address{}

\end{frontmatter}

\section{Broad diameter distribution sample (Sample 1)}

\begin{figure*}[ht]
	\centering
	\includegraphics[scale=0.8]{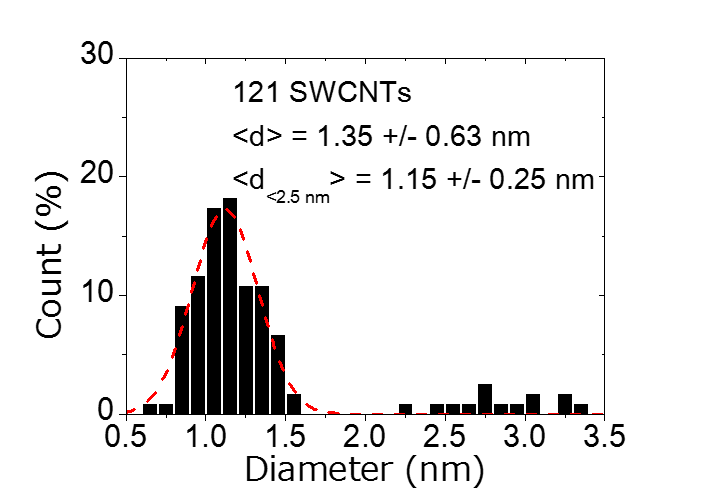}
	\caption{TEM-extracted diameter distribution histogram for Sample 1 (black columns), and Gaussian fit (centered at 1.12 nm, with a full width at half maximum of 0.48 nm). When excluding the SWCNTs with diameters above 2.5 nm (which cannot be detected with the used Raman setup), the mean diameter is 1.15 nm, with a standard deviation of the width of the distribution of 0.25 nm.}
	\label{TEM-dt-sorted}
\end{figure*}

\begin{figure*}[ht]
	\centering
	\includegraphics[scale=1.5]{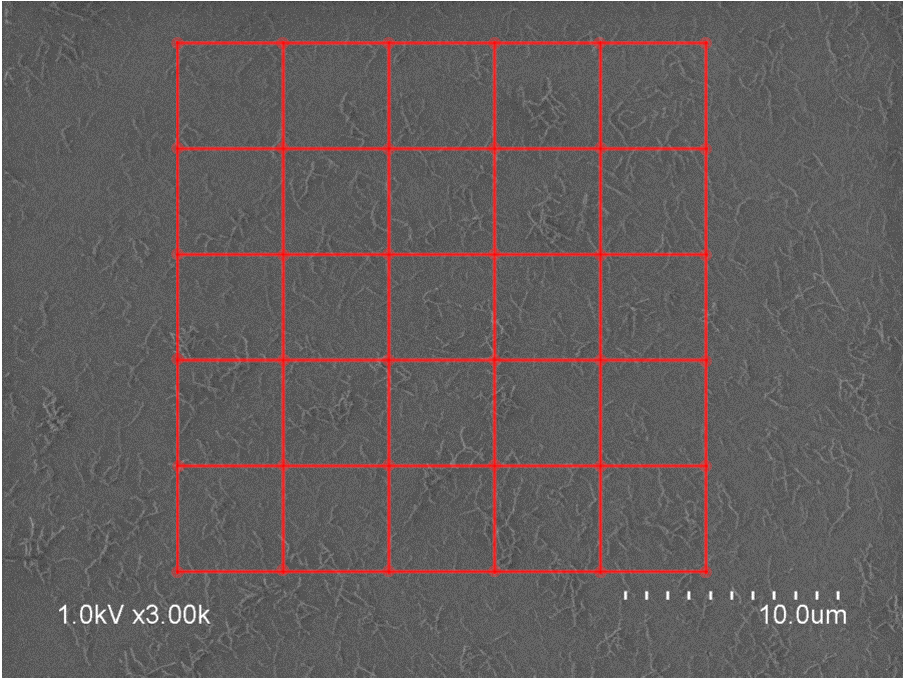}
	\caption{Typical scanning electron microscopy (SEM) image of the broad diameter distribution SWCNTs (Sample 1) deposited onto a SiO$_2$/Si wafer. The red grid highlights the nodes of the Raman peak counting methodology, spaced by 5 $\mu$m. We can clearly see that the step size exceeds the length of the observed SWCNTs.}
	\label{SEM-dt-sorted}
\end{figure*}
\clearpage
\section{Influence of the RBM frequency-to-diameter law}

\begin{figure*}[h!]
	\centering
	\includegraphics[scale=0.28]{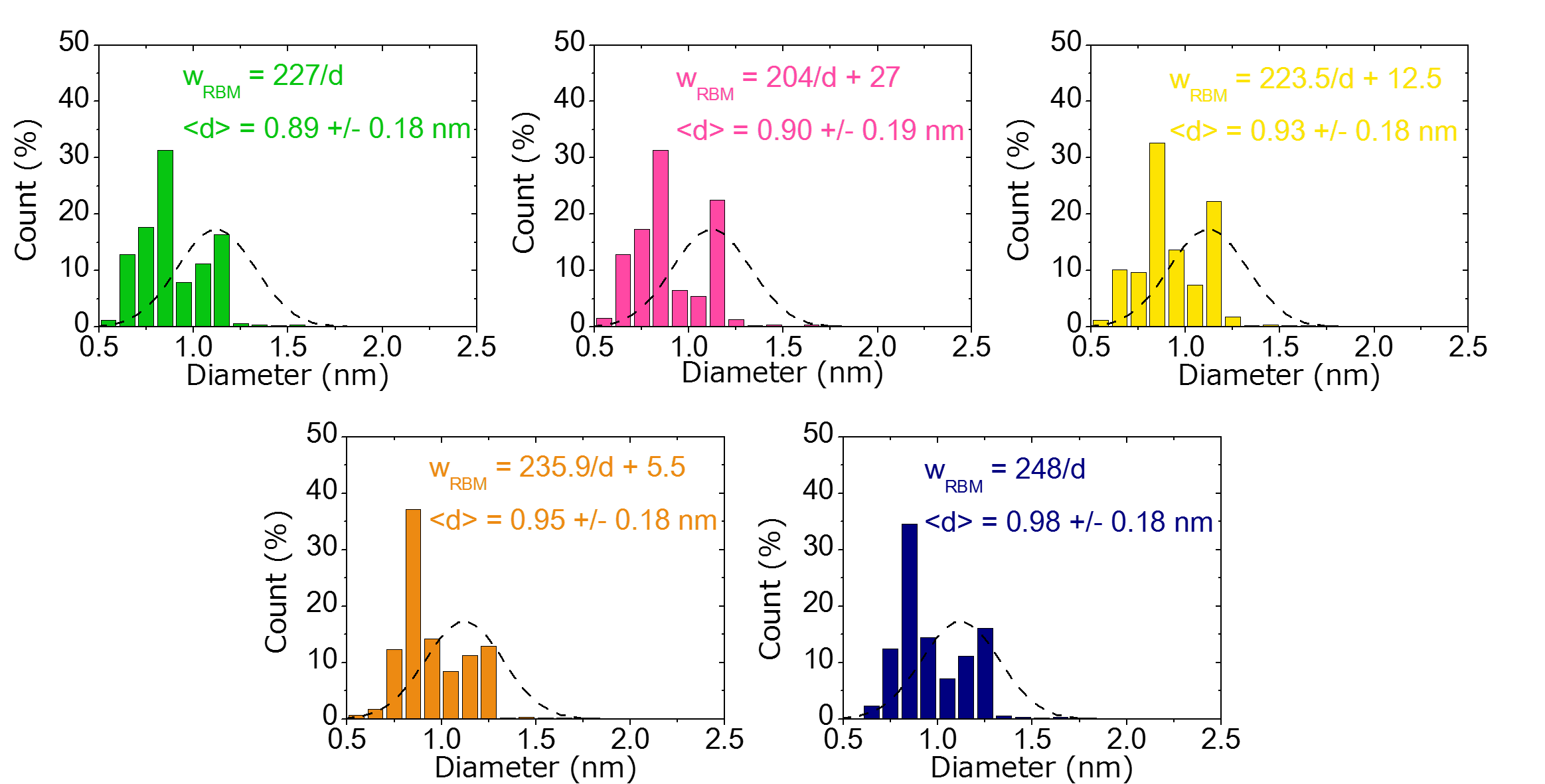}
	\caption{Raman-extracted diameter distribution histograms for the diameter-sorted sample using various RBM frequency-to-diameter laws, along with the Gaussian fit of the TEM-extracted diameter distribution (black dashed line). Clockwise: Ref.~\citenum{araujo2008nature, paillet2006raman, bachilo2002structure, jorio2001structural, zhang2015n}. For each distribution, the mean diameter and corresponding standard deviation are given.}
	\label{Raman-RBM-dt-law}
\end{figure*}

To evaluate the influence of the RBM frequency-to-diameter law on the resulting Raman-extracted diameter distribution, we compared 5 different empirical laws available in the literature to the one used in the main text. Some of the relations are not adapted to the sample type, and/or have been shown to not be entirely accurate. However, comparing the resulting diameter distribution histograms shown on Figure \ref{Raman-RBM-dt-law}, it can be easily established that the observed differences are not significant enough to account for the differences between the TEM- and Raman-extracted distributions shown in Figure 1 in the main text. Though some shifts can be observed, the general shape of these distributions are very similar, as expected.  
\clearpage
\section{Influence of substrate thickness}
\begin{figure*}[h!]
	\centering
	\includegraphics[scale=0.6]{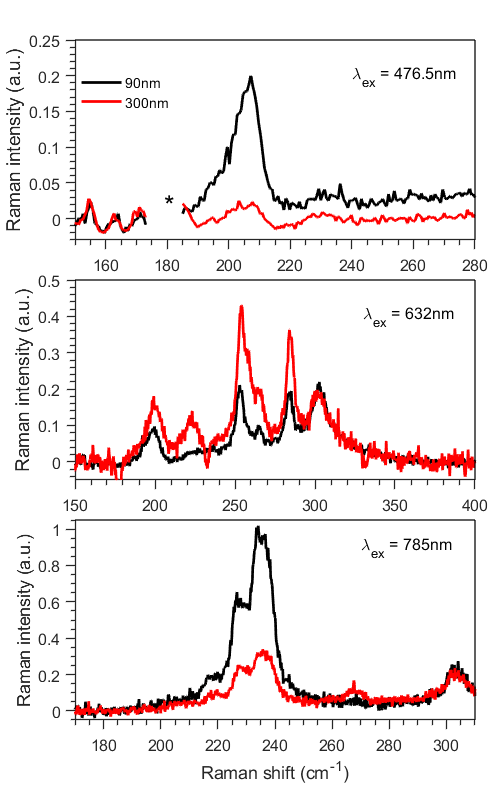}
	\caption{Macroscopic Raman spectra at 3 different excitation wavelengths for samples dropcasted on two substrates with different oxide-layer thicknesses (90 nm and 300 nm). Raman spectra were obtained within a macroscopic setup to excite multiple CNTs at once and were collected at the exact same sample spot. All spectra are normalized using the silicon signal at 520 cm$^{-1}$ from a bare Si sample. The sample preparation does not allow for a direct comparison of RBM intensities between the two samples at a given wavelength, but it is clear that when using the 90 nm substrate, the signal intensities in the blue and IR are much less suppressed, allowing to observe these chiralities in the micro-imaging.}
	\label{SiO2thickness}
\end{figure*}

\section{(n,m) assignment from HRTEM images}

\begin{figure*}[ht!]
	\centering
	\includegraphics[scale=0.5]{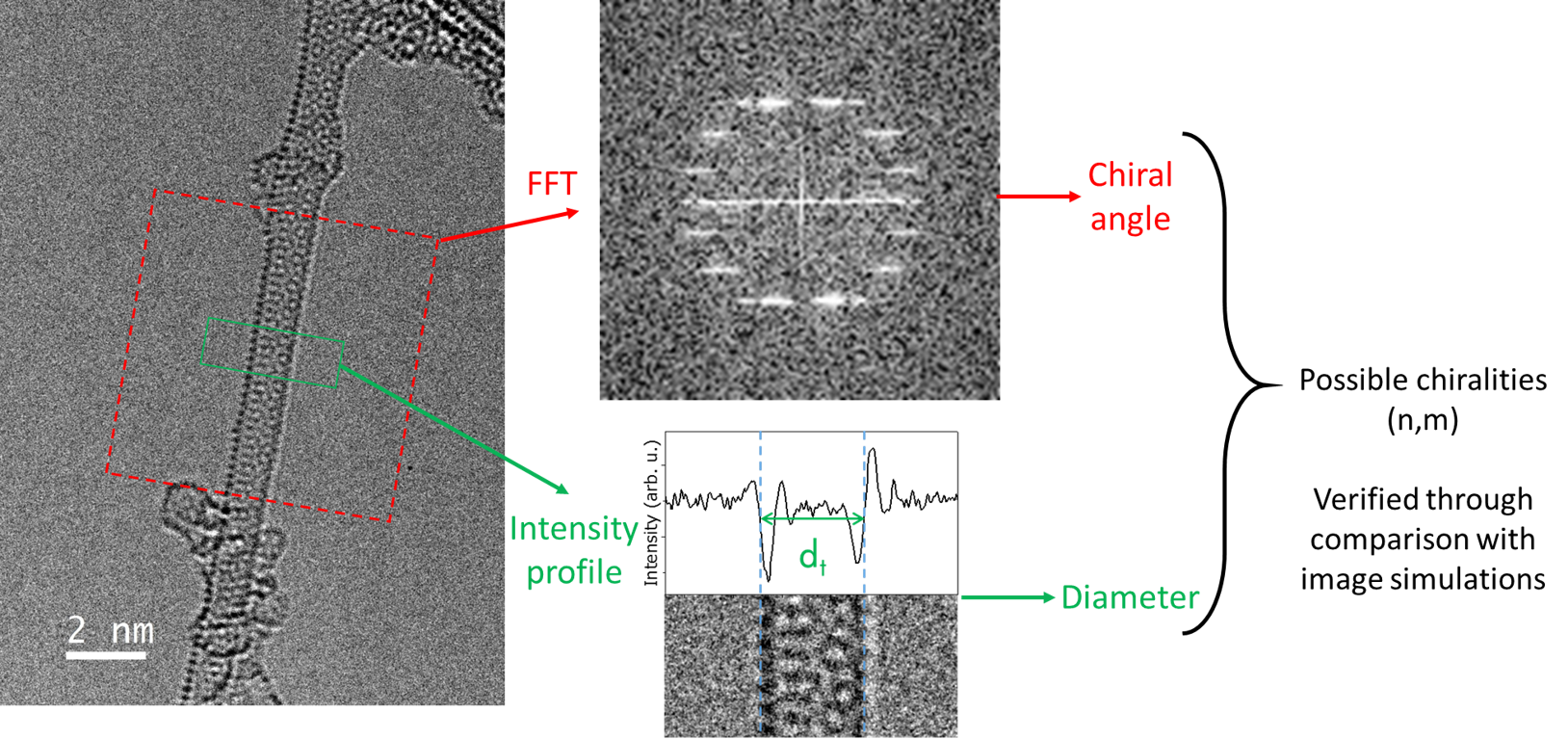}
	\caption{Presentation of the methodology for (n,m) assignments from HRTEM images.}
	\label{HRTEM-method}
\end{figure*}

To determine the chiralities of SWCNTs based on their HRTEM images, the following procedure was used (see Figure \ref{HRTEM-method}). First, the diameter of the nanotube is measured by extracting a contrast intensity profile perpendicular to the tube axis. According to image simulations, the diameter at Scherzer defocus is equal to the distance between the two inflection points of the derivative of the contrast intensity, meaning the midpoint between the minimum of the dark fringe, and the maximum of the subsequent light fringe.\\\
A Fourier transform (FT) of the HRTEM image is then used to determine the chiral angle ($\theta$) of the nanotube. The diffraction pattern of a SWCNT is composed of two main signals. The first are (hk0) reflections from the overlapping lower and upper SWCNT parts, located at the vertices of two concentric hexagons, separated by a rotation angle equal to $2\theta$. The second characteristic signals are intensity maxima along direction lines perpendicular to the tube axis, which correspond to the FT of the cylindrical form factor of the SWCNT, which are Bessel functions. Those lines are called layerlines, and the one passing through the hexagon center (origin of the reciprocal space) is the equatorial line. The chiral angle can be precisely determined by measuring distances between
certain layerlines. Considering d$_1$, d$_2$, and d$_3$ the distances between the equatorial line and the first, second, and third layerlines, respectively, the chiral angle is given by \cite{deniz2010systematic}:
\begin{equation}
\theta = \arctan(\frac{2d_2 - d_3}{\sqrt{3}d_3})
\label{eqn:thetadiff}
\end{equation}
Considering the measured diameter and chiral angle values (with error bars), we can assign several possible (n,m) pairs for the nanotube. Comparing the FT with a calculated diffraction pattern for these chiralities can help take out some possibilities. To discriminate between the possible (n,m) pairs, the HRTEM image is compared to image simulations of all the possible chiralities. The comparison of the Moir\'e patterns leads to an unambiguous (n,m) assignment \cite{ghedjatti2017structural}. Simulated HRTEM images were calculated within dynamical theory with a commercial code (JEMS \cite{stadelmann1987ems}) and homemade software (DiffractX \cite{kirkland2010advanced, kociak2003accurate}).
\clearpage
\section{Details of sample compositions from optical techniques and HRTEM (Samples 2 and 3)}
\begin{figure*}[h!]
	\centering
	\includegraphics[scale=0.8]{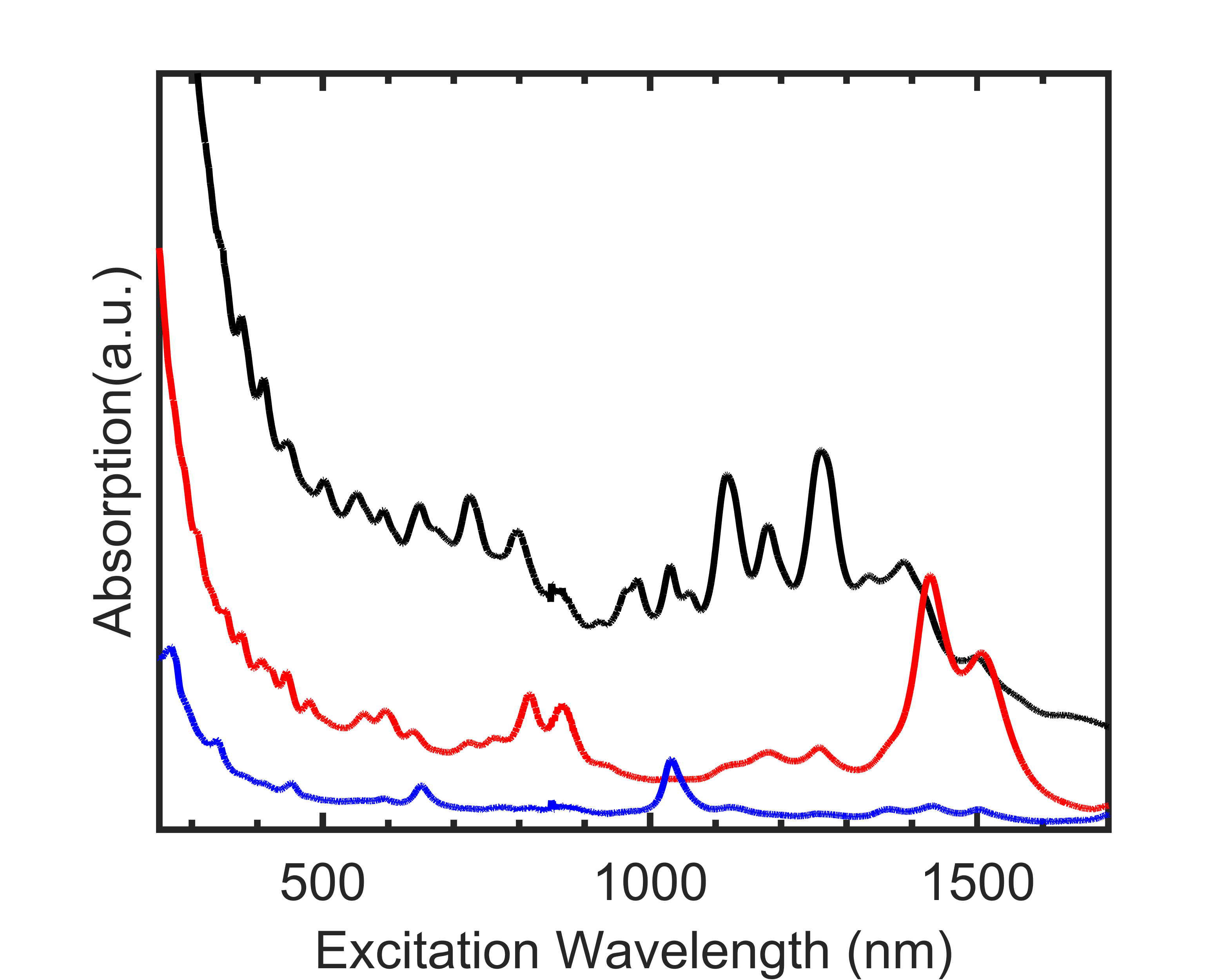}
	\caption{Optical absorption spectra for the parent HiPco sample (black line), Sample 2 (blue line), and Sample 3 (red line).}
	\label{PLE-maps}
\end{figure*}
\begin{figure*}[h!]
	\centering
	\includegraphics[scale=0.8]{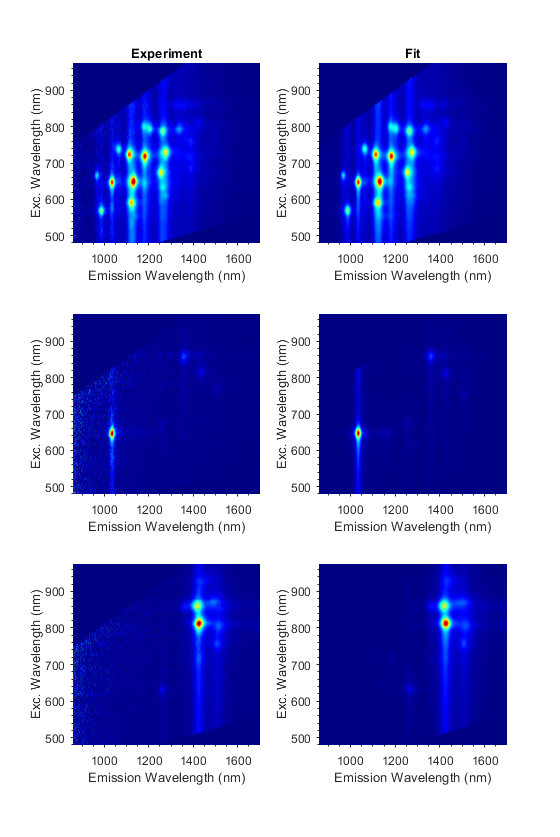}
	\caption{Experimental PLE maps (left), and corresponding fits (right) for, from top to bottom, the parent HiPco sample, Sample 2, and Sample 3. For a more detailed discussion on the fitting program, we refer to the supporting information of reference \cite{li2019separation}}
	\label{PLE-maps}
\end{figure*}
\begin{figure*}[h!]
	\centering
	\includegraphics[scale=0.8]{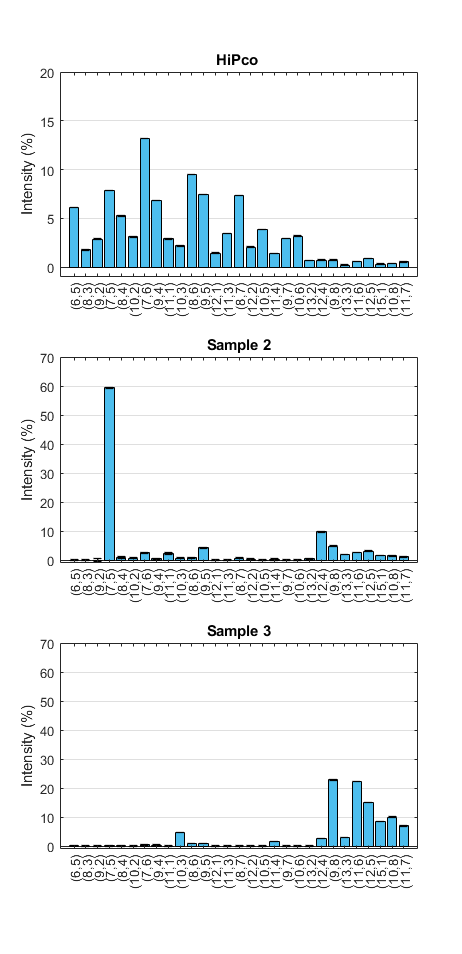}
	\caption{Fitted PLE amplitudes for the parent HiPco sample, Sample 2, and Sample 3. PL intensities were extracted similarly as in reference \cite{li2019separation}}
	\label{PLE-fitted-amplitudes}
\end{figure*}
\clearpage

\begin{table}[]
\centering
\begin{tabular}{lllllllll}
\multicolumn{9}{l}{HRTEM (n,m) assignment results}                                                                                                                                                                                                       \\ \hline
                               &                                & \multicolumn{1}{l|}{}         & \multicolumn{3}{l|}{Number of SWCNTs}                                                & \multicolumn{3}{l}{\% of SWCNTs}                               \\ \hline
\multicolumn{1}{l|}{Chirality} & \multicolumn{1}{l|}{Metallic?} & \multicolumn{1}{l|}{Diameter} & \multicolumn{1}{l|}{HiPco} & \multicolumn{1}{l|}{Sample 2} & \multicolumn{1}{l|}{Sample 3} & \multicolumn{1}{l|}{HiPco} & \multicolumn{1}{l|}{Sample 2} & Sample 3 \\ \hline
\multicolumn{1}{l|}{(9,1)}    & \multicolumn{1}{l|}{0}         & \multicolumn{1}{l|}{0.747}    & \multicolumn{1}{l|}{0}     & \multicolumn{1}{l|}{1}     & \multicolumn{1}{l|}{0}     & \multicolumn{1}{l|}{0.00}  & \multicolumn{1}{l|}{5.00}  & 0.00  \\
\multicolumn{1}{l|}{(7,5)}     & \multicolumn{1}{l|}{0}         & \multicolumn{1}{l|}{0.817}    & \multicolumn{1}{l|}{0}     & \multicolumn{1}{l|}{7}     & \multicolumn{1}{l|}{0}     & \multicolumn{1}{l|}{0.00}  & \multicolumn{1}{l|}{35.00} & 0.00  \\
\multicolumn{1}{l|}{(8,4)}     & \multicolumn{1}{l|}{0}         & \multicolumn{1}{l|}{0.829}    & \multicolumn{1}{l|}{2}     & \multicolumn{1}{l|}{0}     & \multicolumn{1}{l|}{0}     & \multicolumn{1}{l|}{7.14}  & \multicolumn{1}{l|}{0.00}  & 0.00  \\
\multicolumn{1}{l|}{(11,0)}    & \multicolumn{1}{l|}{0}         & \multicolumn{1}{l|}{0.861}    & \multicolumn{1}{l|}{1}     & \multicolumn{1}{l|}{0}     & \multicolumn{1}{l|}{0}     & \multicolumn{1}{l|}{3.57}  & \multicolumn{1}{l|}{0.00}  & 0.00  \\
\multicolumn{1}{l|}{(7,6)}     & \multicolumn{1}{l|}{0}         & \multicolumn{1}{l|}{0.882}    & \multicolumn{1}{l|}{3}     & \multicolumn{1}{l|}{0}     & \multicolumn{1}{l|}{0}     & \multicolumn{1}{l|}{10.71} & \multicolumn{1}{l|}{0.00}  & 0.00  \\
\multicolumn{1}{l|}{(8,5)}     & \multicolumn{1}{l|}{1}         & \multicolumn{1}{l|}{0.889}    & \multicolumn{1}{l|}{1}     & \multicolumn{1}{l|}{0}     & \multicolumn{1}{l|}{0}     & \multicolumn{1}{l|}{3.57}  & \multicolumn{1}{l|}{0.00}  & 0.00  \\
\multicolumn{1}{l|}{(9,4)}     & \multicolumn{1}{l|}{0}         & \multicolumn{1}{l|}{0.903}    & \multicolumn{1}{l|}{1}     & \multicolumn{1}{l|}{0}     & \multicolumn{1}{l|}{0}     & \multicolumn{1}{l|}{3.57}  & \multicolumn{1}{l|}{0.00}  & 0.00  \\
\multicolumn{1}{l|}{(11,1)}    & \multicolumn{1}{l|}{0}         & \multicolumn{1}{l|}{0.903}    & \multicolumn{1}{l|}{0}     & \multicolumn{1}{l|}{1}     & \multicolumn{1}{l|}{0}     & \multicolumn{1}{l|}{0.00}  & \multicolumn{1}{l|}{5.00}  & 0.00  \\
\multicolumn{1}{l|}{(10,3)}    & \multicolumn{1}{l|}{0}         & \multicolumn{1}{l|}{0.923}    & \multicolumn{1}{l|}{0}     & \multicolumn{1}{l|}{0}     & \multicolumn{1}{l|}{1}     & \multicolumn{1}{l|}{0.00}  & \multicolumn{1}{l|}{0.00}  & 1.82  \\
\multicolumn{1}{l|}{(12,0)}    & \multicolumn{1}{l|}{1}         & \multicolumn{1}{l|}{0.939}    & \multicolumn{1}{l|}{0}     & \multicolumn{1}{l|}{0}     & \multicolumn{1}{l|}{1}     & \multicolumn{1}{l|}{0.00}  & \multicolumn{1}{l|}{0.00}  & 1.82  \\
\multicolumn{1}{l|}{(8,6)}     & \multicolumn{1}{l|}{0}         & \multicolumn{1}{l|}{0.952}    & \multicolumn{1}{l|}{2}     & \multicolumn{1}{l|}{0}     & \multicolumn{1}{l|}{1}     & \multicolumn{1}{l|}{7.14}  & \multicolumn{1}{l|}{0.00}  & 1.82  \\
\multicolumn{1}{l|}{(9,5)}     & \multicolumn{1}{l|}{0}         & \multicolumn{1}{l|}{0.962}    & \multicolumn{1}{l|}{1}     & \multicolumn{1}{l|}{1}     & \multicolumn{1}{l|}{0}     & \multicolumn{1}{l|}{3.57}  & \multicolumn{1}{l|}{5.00}  & 0.00  \\
\multicolumn{1}{l|}{(12,1)}    & \multicolumn{1}{l|}{0}         & \multicolumn{1}{l|}{0.981}    & \multicolumn{1}{l|}{1}     & \multicolumn{1}{l|}{1}     & \multicolumn{1}{l|}{0}     & \multicolumn{1}{l|}{3.57}  & \multicolumn{1}{l|}{5.00}  & 0.00  \\
\multicolumn{1}{l|}{(11,3)}    & \multicolumn{1}{l|}{0}         & \multicolumn{1}{l|}{1.000}    & \multicolumn{1}{l|}{1}     & \multicolumn{1}{l|}{0}     & \multicolumn{1}{l|}{1}     & \multicolumn{1}{l|}{3.57}  & \multicolumn{1}{l|}{0.00}  & 1.82  \\
\multicolumn{1}{l|}{(8,7)}     & \multicolumn{1}{l|}{0}         & \multicolumn{1}{l|}{1.018}    & \multicolumn{1}{l|}{2}     & \multicolumn{1}{l|}{0}     & \multicolumn{1}{l|}{0}     & \multicolumn{1}{l|}{7.14}  & \multicolumn{1}{l|}{0.00}  & 0.00  \\
\multicolumn{1}{l|}{(12,2)}    & \multicolumn{1}{l|}{0}         & \multicolumn{1}{l|}{1.027}    & \multicolumn{1}{l|}{1}     & \multicolumn{1}{l|}{0}     & \multicolumn{1}{l|}{0}     & \multicolumn{1}{l|}{3.57}  & \multicolumn{1}{l|}{0.00}  & 0.00  \\
\multicolumn{1}{l|}{(10,5)}    & \multicolumn{1}{l|}{0}         & \multicolumn{1}{l|}{1.036}    & \multicolumn{1}{l|}{2}     & \multicolumn{1}{l|}{1}     & \multicolumn{1}{l|}{0}     & \multicolumn{1}{l|}{7.14}  & \multicolumn{1}{l|}{5.00}  & 0.00  \\
\multicolumn{1}{l|}{(13,1)}    & \multicolumn{1}{l|}{1}         & \multicolumn{1}{l|}{1.059}    & \multicolumn{1}{l|}{1}     & \multicolumn{1}{l|}{0}     & \multicolumn{1}{l|}{0}     & \multicolumn{1}{l|}{3.57}  & \multicolumn{1}{l|}{0.00}  & 0.00  \\
\multicolumn{1}{l|}{(9,7)}     & \multicolumn{1}{l|}{0}         & \multicolumn{1}{l|}{1.088}    & \multicolumn{1}{l|}{1}     & \multicolumn{1}{l|}{1}     & \multicolumn{1}{l|}{2}     & \multicolumn{1}{l|}{3.57}  & \multicolumn{1}{l|}{5.00}  & 3.64  \\
\multicolumn{1}{l|}{(10,6)}    & \multicolumn{1}{l|}{0}         & \multicolumn{1}{l|}{1.096}    & \multicolumn{1}{l|}{0}     & \multicolumn{1}{l|}{0}     & \multicolumn{1}{l|}{1}     & \multicolumn{1}{l|}{0.00}  & \multicolumn{1}{l|}{0.00}  & 1.82  \\
\multicolumn{1}{l|}{(11,5)}    & \multicolumn{1}{l|}{1}         & \multicolumn{1}{l|}{1.110}    & \multicolumn{1}{l|}{1}     & \multicolumn{1}{l|}{0}     & \multicolumn{1}{l|}{1}     & \multicolumn{1}{l|}{3.57}  & \multicolumn{1}{l|}{0.00}  & 1.82  \\
\multicolumn{1}{l|}{(12,4)}    & \multicolumn{1}{l|}{0}         & \multicolumn{1}{l|}{1.129}    & \multicolumn{1}{l|}{0}     & \multicolumn{1}{l|}{1}     & \multicolumn{1}{l|}{1}     & \multicolumn{1}{l|}{0.00}  & \multicolumn{1}{l|}{5.00}  & 1.82  \\
\multicolumn{1}{l|}{(14,1)}    & \multicolumn{1}{l|}{0}         & \multicolumn{1}{l|}{1.137}    & \multicolumn{1}{l|}{0}     & \multicolumn{1}{l|}{0}     & \multicolumn{1}{l|}{3}     & \multicolumn{1}{l|}{0.00}  & \multicolumn{1}{l|}{0.00}  & 5.45  \\
\multicolumn{1}{l|}{(9,8)}     & \multicolumn{1}{l|}{0}         & \multicolumn{1}{l|}{1.153}    & \multicolumn{1}{l|}{1}     & \multicolumn{1}{l|}{3}     & \multicolumn{1}{l|}{15}    & \multicolumn{1}{l|}{3.57}  & \multicolumn{1}{l|}{15.00} & 27.27 \\
\multicolumn{1}{l|}{(13,3)}    & \multicolumn{1}{l|}{0}         & \multicolumn{1}{l|}{1.153}    & \multicolumn{1}{l|}{0}     & \multicolumn{1}{l|}{1}     & \multicolumn{1}{l|}{4}     & \multicolumn{1}{l|}{0.00}  & \multicolumn{1}{l|}{5.00}  & 7.27  \\
\multicolumn{1}{l|}{(11,6)}    & \multicolumn{1}{l|}{0}         & \multicolumn{1}{l|}{1.169}    & \multicolumn{1}{l|}{0}     & \multicolumn{1}{l|}{1}     & \multicolumn{1}{l|}{3}     & \multicolumn{1}{l|}{0.00}  & \multicolumn{1}{l|}{5.00}  & 5.45  \\
\multicolumn{1}{l|}{(14,2)}    & \multicolumn{1}{l|}{1}         & \multicolumn{1}{l|}{1.182}    & \multicolumn{1}{l|}{0}     & \multicolumn{1}{l|}{0}     & \multicolumn{1}{l|}{2}     & \multicolumn{1}{l|}{0.00}  & \multicolumn{1}{l|}{0.00}  & 3.64  \\
\multicolumn{1}{l|}{(12,5)}    & \multicolumn{1}{l|}{0}         & \multicolumn{1}{l|}{1.185}    & \multicolumn{1}{l|}{0}     & \multicolumn{1}{l|}{0}     & \multicolumn{1}{l|}{4}     & \multicolumn{1}{l|}{0.00}  & \multicolumn{1}{l|}{0.00}  & 7.27  \\
\multicolumn{1}{l|}{(15,1)}    & \multicolumn{1}{l|}{0}         & \multicolumn{1}{l|}{1.215}    & \multicolumn{1}{l|}{0}     & \multicolumn{1}{l|}{1}     & \multicolumn{1}{l|}{1}     & \multicolumn{1}{l|}{0.00}  & \multicolumn{1}{l|}{5.00}  & 1.82  \\
\multicolumn{1}{l|}{(9,9)}     & \multicolumn{1}{l|}{1}         & \multicolumn{1}{l|}{1.220}    & \multicolumn{1}{l|}{0}     & \multicolumn{1}{l|}{0}     & \multicolumn{1}{l|}{4}     & \multicolumn{1}{l|}{0.00}  & \multicolumn{1}{l|}{0.00}  & 7.27  \\
\multicolumn{1}{l|}{(10,8)}    & \multicolumn{1}{l|}{0}         & \multicolumn{1}{l|}{1.223}    & \multicolumn{1}{l|}{1}     & \multicolumn{1}{l|}{0}     & \multicolumn{1}{l|}{4}     & \multicolumn{1}{l|}{3.57}  & \multicolumn{1}{l|}{0.00}  & 7.27  \\
\multicolumn{1}{l|}{(11,7)}    & \multicolumn{1}{l|}{0}         & \multicolumn{1}{l|}{1.230}    & \multicolumn{1}{l|}{1}     & \multicolumn{1}{l|}{0}     & \multicolumn{1}{l|}{5}     & \multicolumn{1}{l|}{3.57}  & \multicolumn{1}{l|}{0.00}  & 9.09  \\
\multicolumn{1}{l|}{(12,6)}    & \multicolumn{1}{l|}{1}         & \multicolumn{1}{l|}{1.243}    & \multicolumn{1}{l|}{2}     & \multicolumn{1}{l|}{0}     & \multicolumn{1}{l|}{0}     & \multicolumn{1}{l|}{7.14}  & \multicolumn{1}{l|}{0.00}  & 0.00  \\
\multicolumn{1}{l|}{(13,5)}    & \multicolumn{1}{l|}{0}         & \multicolumn{1}{l|}{1.260}    & \multicolumn{1}{l|}{1}     & \multicolumn{1}{l|}{0}     & \multicolumn{1}{l|}{1}     & \multicolumn{1}{l|}{3.57}  & \multicolumn{1}{l|}{0.00}  & 1.82  \\
\multicolumn{1}{l|}{(13,7)}    & \multicolumn{1}{l|}{1}         & \multicolumn{1}{l|}{1.376}    & \multicolumn{1}{l|}{1}     & \multicolumn{1}{l|}{0}     & \multicolumn{1}{l|}{0}     & \multicolumn{1}{l|}{3.57}  & \multicolumn{1}{l|}{0.00}  & 0.00 
\end{tabular}
\caption{Results for the (n,m) assignments from the HRTEM characterization of the parent HiPco sample and the two chirality-enriched samples. The diameters are calculated using a 1.42 \AA\ carbon-carbon bond length.}
\end{table}

\clearpage

\section{Raman peak counting and wavelength-dependent Raman for chirality-enriched samples (Samples 2 and 3)}
\begin{figure*}[h!]
	\centering
	\includegraphics[scale=0.53]{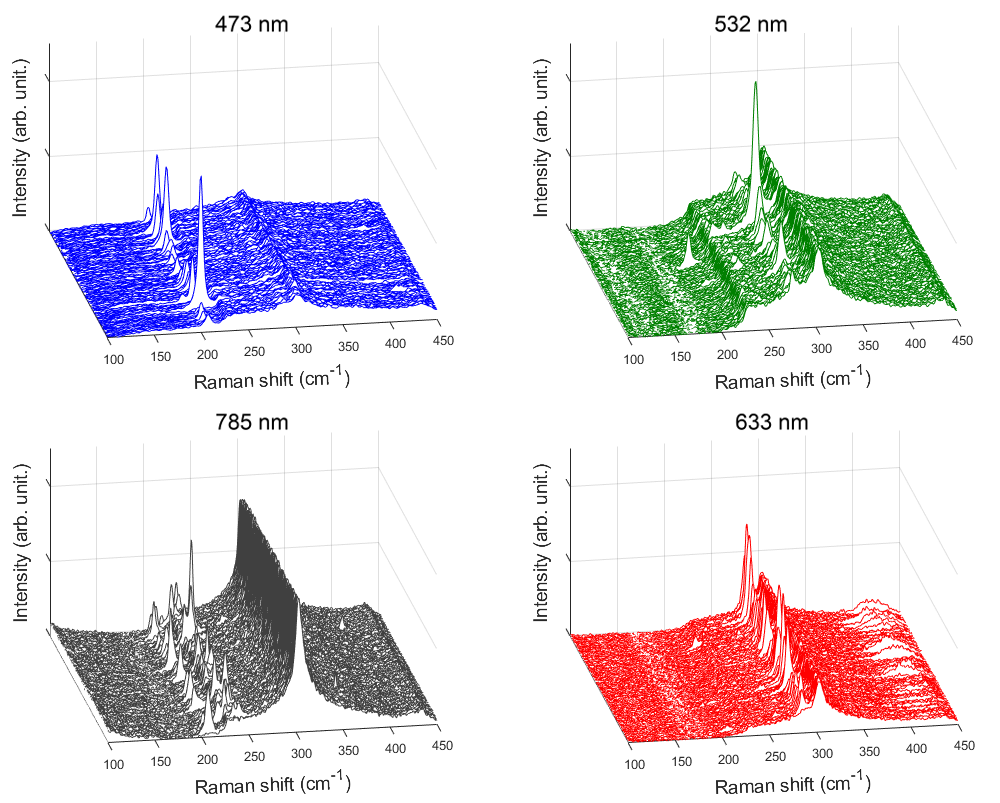}
	\caption{Examples of Raman mappings used for RBM peak counting acquired on a flat substrate in the RBM spectral range at each excitation wavelength for Sample 2. The peak at 303 cm$^{-1}$ is attributed to the SiO$_2$ substrate.}
	\label{RBM-flat-S2}
\end{figure*}
\clearpage
\begin{figure*}[h!]
	\centering
	\includegraphics[scale=0.53]{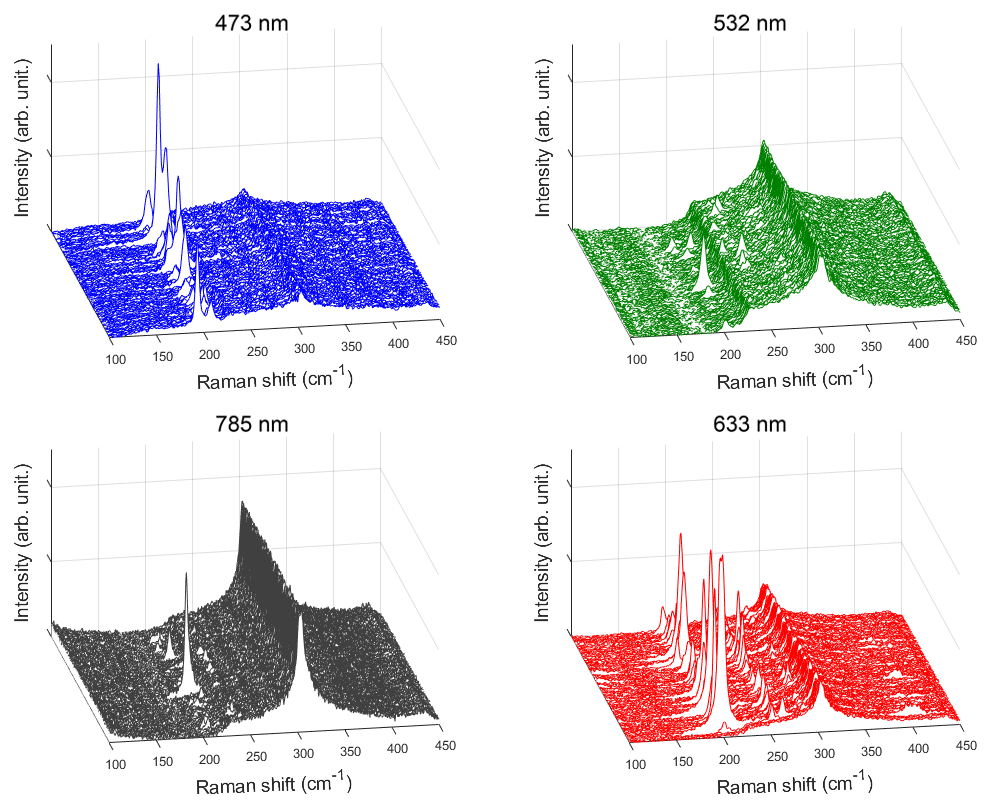}
	\caption{Examples of Raman mappings used for RBM peak counting acquired on a flat substrate in the RBM spectral range at each excitation wavelength for Sample 3. The peak at 303 cm$^{-1}$ is attributed to the SiO$_2$ substrate.}
	\label{RBM-flat-S3}
\end{figure*}
\begin{figure*}[h!]
	\centering
	\includegraphics[scale=0.06]{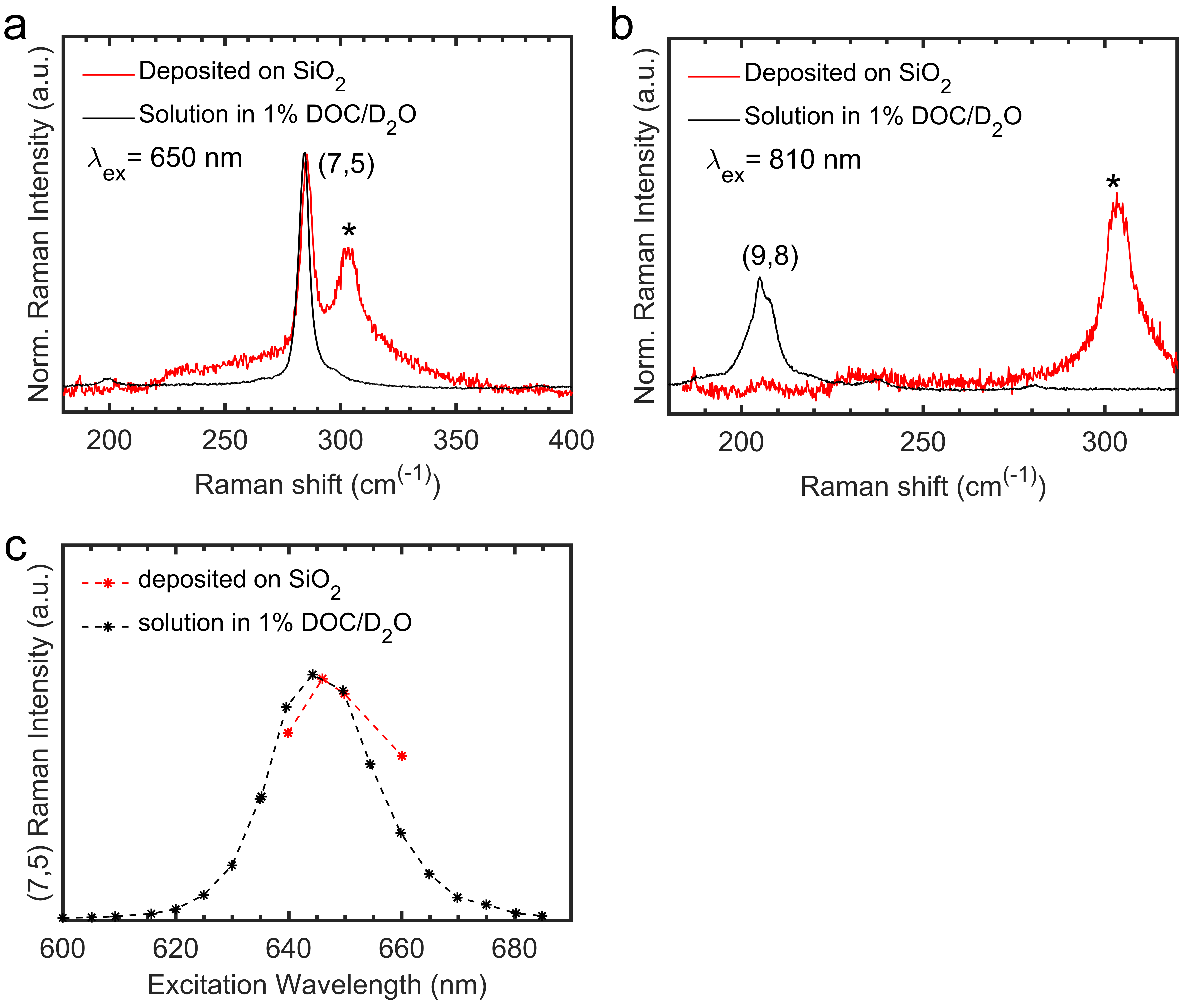}
	\caption{Comparison of Raman results on flat substrates and in solution. Raman spectra acquired on SiO$_2$/Si wafers (same samples used for Raman peak counting) (red line) and in solution in 1$\%$ DOC/D$_2$O (black line) for a) Sample 2, and b) Sample 3. For Sample 2, spectra were acquired at an excitation wavelength of 650 nm, to fall into the resonance window of the (7,5) SWCNT, and for Sample 3, spectra were acquired at 810 nm to fall into the resonance window of the (9,8) SWCNT. In the case of the (7,5) chirality, the RBM frequency is nearly the same in both environments. The peaks denoted with an asterisk are attributed to the SiO$_2$ substrate. c) Raman resonance profiles for the (7,5) SWCNT measured in Sample 2 on a  SiO$_2$/Si substrate (red), and in solution (black). }
	\label{RBM-flat-98}
\end{figure*}
\begin{figure*}[h!]
	\centering
	\includegraphics[scale=0.75]{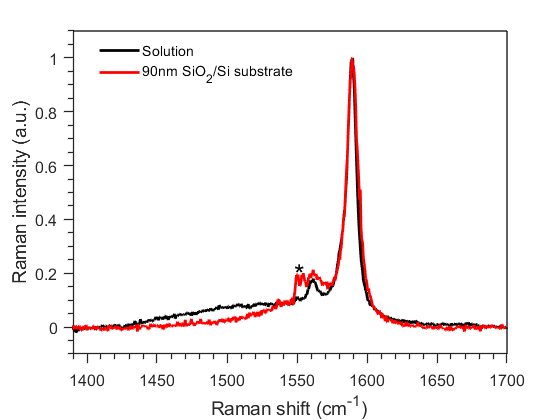}
	\caption{Raman spectra of the HiPco SWCNT ‘parent’ sample in solution (black) or deposited on a 90 nm SiO$_2$/Si substrate acquired in a macroscopic Raman setup excited at 476.5 nm of an Ar+ ion laser. No shift can be observed for the G-band modes due to doping of the substrate. The * denotes the presence of a plasma line of the laser, which is only observed for the substrate sample due to the direct reflection of the laser on the substrate.}
	\label{G-band}
\end{figure*}
\begin{figure*}[h!]
	\centering
	\includegraphics[scale=0.75]{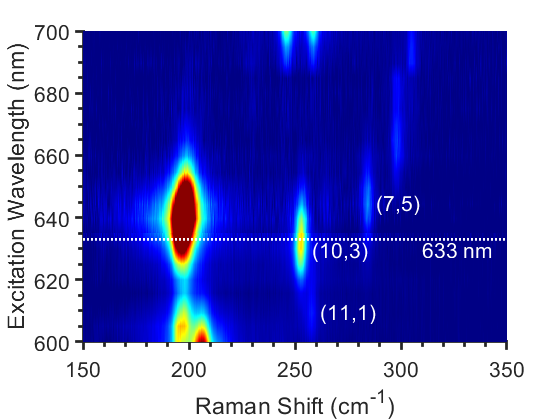}
	\caption{Wavelength-dependent Raman map with 5 nm step size obtained in liquid suspension for Sample 3, zoomed-in on the region of the 633 nm laser excitation.}
	\label{10-3}
\end{figure*}
\clearpage

\bibliographystyle{elsarticle-num} 
\bibliography{Article-AC-TEM-Raman-sf}